\documentclass{aa}  

\usepackage{graphicx}
\usepackage{txfonts}
\usepackage{lipsum}
\usepackage{subcaption}
\usepackage{lscape}
\usepackage{placeins}
\usepackage[colorlinks = true, allcolors=blue]{hyperref}
\usepackage[normalem]{ulem}
\usepackage{orcidlink}
\usepackage{booktabs}

\usepackage{graphicx}
\usepackage{txfonts}
\usepackage{subcaption}         % necessary for continued figures, example in section 3
\usepackage{lscape}             % to rotate a single page table, example in appendix.
\usepackage{placeins}           % useful with 

\newcommand{\Ms}{\ensuremath{{\rm~M}_{\odot}}}
\newcommand{\Zs}{\ensuremath{{\rm~Z}_{\odot}}}

\newcommand{\ie}{{\it i.e.}}

\newcommand{\nbseven}{{\tt NBODY7}}

\newcommand{\fbin}{\ensuremath{f_{\rm bin}}}
\newcommand{\mcl}{\ensuremath{M_{\rm cl}}}
\newcommand{\rh}{\ensuremath{r_{\rm h}}}
\newcommand{\mone}{\ensuremath{{\rm M}_1}}
\newcommand{\mtwo}{\ensuremath{{\rm M}_2}}
\newcommand{\mc}{\ensuremath{{\rm M}_{\rm C}}}
\newcommand{\nmrg}{\ensuremath{N_{\rm mrg}}}
\newcommand{\etamrg}{\ensuremath{\eta_{\rm mrg}}}

\begin{document}

\title{Examining the stellar-merger origin of the blue main sequence in the open cluster NGC\,3532 with N-body simulations}

\titlerunning{N-body simulation of NGC 3532 open cluster}

\author{
Khushboo K. Rao\orcidlink{0000-0001-7470-9192}\inst{1}\corrauth{khushboo@astro.ncu.edu.tw}
\and 
Sambaran Banerjee\orcidlink{0000-0002-1254-2603}\inst{2}
}
\institute{
Institute of Astronomy, National Central University, 300 Zhongda Road, Zhongli 32001 Taoyuan, Taiwan
\and
Helmholtz-Instituts für Strahlen- und Kernphysik (HISKP), Nussallee 14-16, D-53115 Bonn, Germany
}

%%%%%%%%%%%%%%%%%%%%%%%%%%%%%%%%%%%%%%%%%%%%%%%%%%
% Abstract
  \abstract
    {Extended main-sequence turnoffs, extended main sequences, and split main sequences observed in the colour-magnitude diagrams of young and intermediate-age star clusters are now widely interpreted as the consequence of a distribution of stellar rotation rates among their intermediate-mass (.5--1.8~M$_\odot$) members. However, the origin of the slowly rotating population that occupies the blue main sequence (bMS) remains uncertain, and stellar mergers have been proposed as one possible pathway.} 
    {We investigate whether stellar mergers can account for the observed bMS population in the 330-Myr Galactic open cluster NGC\,3532.} 
    {We perform fourteen direct $\nbseven$ simulations spanning different initial cluster masses, radii, binary fractions, and binary orbital distributions. The simulations are selected to reproduce the present-day properties of NGC\,3532 within the observational uncertainties, allowing us to estimate the expected number of merger products among its intermediate-mass main-sequence (MS) members.} 
    {The cluster hosts $\approx 37\%$ bMS members among its intermediate-mass MS population, which are predominantly slow rotators. Despite this large bMS population, our simulations produce only a handful of MS-MS merger products, due to the cluster's low density and substantial mass loss during its evolution.}
    {Stellar mergers are unlikely to be the dominant formation channel for the observed slowly rotating bMS population in NGC\,3532 and other disperse open clusters. Our results instead favour angular-momentum loss mechanisms operating before or shortly after the zero-age main sequence, such as pre-main-sequence star-disk interactions or tidal synchronization in low-mass ratio binaries.} 
    
    
\maketitle
\nolinenumbers
%%%%%%%%%%%%%%%%% BODY OF PAPER %%%%%%%%%%%%%%%%%%

%--------------Introduction ------------------%

\section{Introduction} \label{sec:intro}
Extended main sequences (eMSs) and extended main sequence turnoffs (eMSTOs) have been widely observed in colour-magnitude diagrams (CMDs) of young and intermediate-age clusters of ages $<2$~Gyr \citep{Milone2009A&A...497..755M, Marino2018ApJ...863L..33M, Deng2024RAA....24f5004D}. In these clusters, MSTO regions or the upper main sequences (MS) appear substantially broader than other evolutionary sequences, such as the lower MS, subgiant branch, and red giant branch. A distribution of stellar rotation rates among intermediate-mass stars (1.5--8~M$_\odot$) is now widely accepted as the primary driver of these features \citep{Bastian2009MNRAS.398L..11B, Li2012ApJ...761L..22L, Bastian2018MNRAS.480.3739B, Correnti2021MNRAS.504..155C}, with fast and slow rotators preferentially populating the redder and bluer sides of MS, respectively \citep{Bastian2018MNRAS.480.3739B, Kamann2020MNRAS.492.2177K}. The slow rotators are identified to have projected equatorial velocities ($v \sin i$) as small as $<10$~km~s$^{-1}$ to $ \approx 100$~km~s$^{-1}$, whereas fast rotators can have $v \sin i > 100$~km~s$^{-1}$ to near-critical rotation approaching $\sim 400$~km~s$^{-1}$ \citep{Bastian2017MNRAS.465.4795B, Kamann2023MNRAS.518.1505K, Cordoni2024MNRAS.532.1547C, Rao2026ApJ..1002..103R}. Recent studies have shown that approximately 40\% of intermediate-mass stars are slow rotators \citep{Kamann2020MNRAS.492.2177K, Correnti2021MNRAS.504..155C, Leanza2025A&A...698A..27L, Rao2026ApJ..1002..103R}. 

While the origin of the rapidly rotating population is reasonably well understood, the formation pathway of the slowly rotating stars remains uncertain. Rapid rotation is a characteristic property of intermediate-mass stars owing to angular momentum conservation during pre-MS contraction en route to the zero-age MS and the presence of radiative envelopes during their MS evolution. Some fast rotators even possess gaseous decretion disks called "Be" or "shell" stars, identified by emission lines in their spectra \citep{Kamann2023MNRAS.518.1505K}. Some of the Be spectral-type stars are also linked to mass transfer from an evolved companion, which feeds the disks of the B-type stars \citep[see][and references therein]{Rivinius2013A&ARv..21...69R}. In contrast, several mechanisms have been proposed to explain the presence of slow rotators, such as tidal synchronization in binaries \citep{DAntona2015MNRAS4532637}, stellar merger \citep{Wang2022NatAs...6..480W}, and star-disk interactions during the pre-MS phase \citep{Bastian2020MNRAS.495.1978B}. However, the extent to which each of these factors contributes to the total fraction of slow rotators remains an open question.

%fig01
\begin{figure}
\begin{center}
    \includegraphics[width=0.95\linewidth]{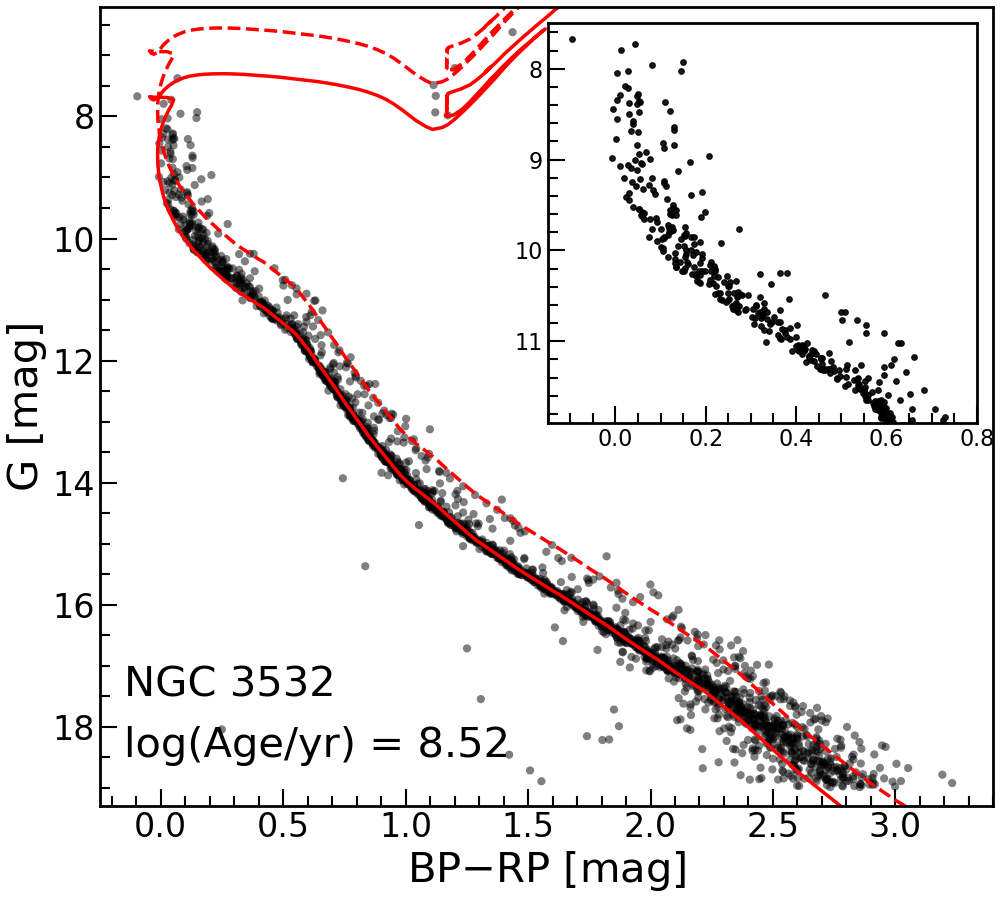}
    \caption{Gaia CMD of NGC\,3532, fitted with a non-rotating isochrone (red solid line) of $\log{\rm (Age/yr)}= 8.52$, distance of 490~pc, $A_{\rm v} = 0.07$, and $[Fe/H] = 0.0$. The red dashed line shows an equal-mass binary track obtained by shifting $G = -0.75$~mag. The inset zooms in on the MSTO region of the cluster, showing a clear eMSTO. }
    \label{fig:obs_CMD}
\end{center}
\end{figure}

In the stellar merger scenario, mergers can occur during mass transfer in binaries \citep{Henneco2024A&A...682A.169H} or due to an outer companion in a hierarchical triple system \citep{Naoz2016ARA&A..54..441N}. Because of angular momentum conservation, newly formed merger products are expected to rotate rapidly, and the final rotation rate depends on the time elapsed since the merger event. Although various mechanisms have been proposed to explain the spin-down of these merged products \citep{Schneider2025arXiv250918421S}, identifying these merger products among ordinary MS stars in CMDs remains extremely challenging. Generally, fast rotation and single-star characteristics are used as proxies for recent stellar merger events. However, because intermediate-mass stars naturally span a broad range of rotation rates, rotation alone cannot reliably distinguish merger products from normal MS stars.

\citet{Wang2020ApJ...888L..12W} investigated the LMC cluster NGC\,1783 to explore the role of stellar mergers in the formation of its observed blue main sequence (bMS). Based on the flat mass function of the bMS compared to the Salpeter mass function of the red main-sequence (rMS) and binary modeling, they proposed that stellar mergers are responsible for the observed bMS in the cluster. Their simulations indicated that most merger events occur early in a cluster's evolution, typically within the first $\sim10$~Myr, with the rate decreasing at later ages, providing sufficient time for the merger products to spin down. However, this interpretation has been challenged. Using $v\sin i$ measurements of intermediate-mass members of NGC\,1783, \citet{Bastian2025A&A...700A.241B} proposed that the bMS is unlikely to be composed primarily of stellar merger products. Instead, they argued that the upper bMS more closely resembles the blue straggler population located in the extension of MSs above MSTOs in CMDs of old globular and open clusters \citep{Sandage1953AJ.....58...61S, Rao2023MNRAS.526.1057R}. As blue stragglers are known to form via different mechanisms, like mass transfer or mergers in binary systems or stellar collisions \citep{Boffin2015ASSL..413.....B}, they have a wide range of rotational velocities \citep{Ferraro2023NatCo..14.2584F}.

Motivated by these contrasting interpretations, we investigate the role of stellar mergers in producing intermediate-mass stars in the Galactic open cluster NGC\,3532. We perform direct N-body simulations to estimate the fraction of merger products that currently populate the intermediate-mass MS. NGC\,3532 is approximately 330~Myr old and is located at a distance of 490~pc \citep{Rao2026ApJ..1002..103R}. The cluster exhibits a prominent eMS with a clear segregation of fast and slow rotators in redder and bluer regions of the CMD \citep{Cordoni2024MNRAS.532.1547C, Rao2025arXiv251205458R, Rao2026ApJ..1002..103R}. Owing to its low foreground extinction ($A_{\rm V}=0.07$~mag), the observed positions of cluster members in the CMD closely represent their intrinsic luminosities and temperatures. These characteristics make NGC\,3532 an excellent laboratory for testing whether stellar mergers can account for the observed population of slowly rotating intermediate-mass stars.

%-------------- Data & Cluster Membership ------------------%

\section{Cluster membership \& properties} \label{sec:data}

%fig02
\begin{figure}
\begin{center}
   \includegraphics[width=0.98\linewidth]{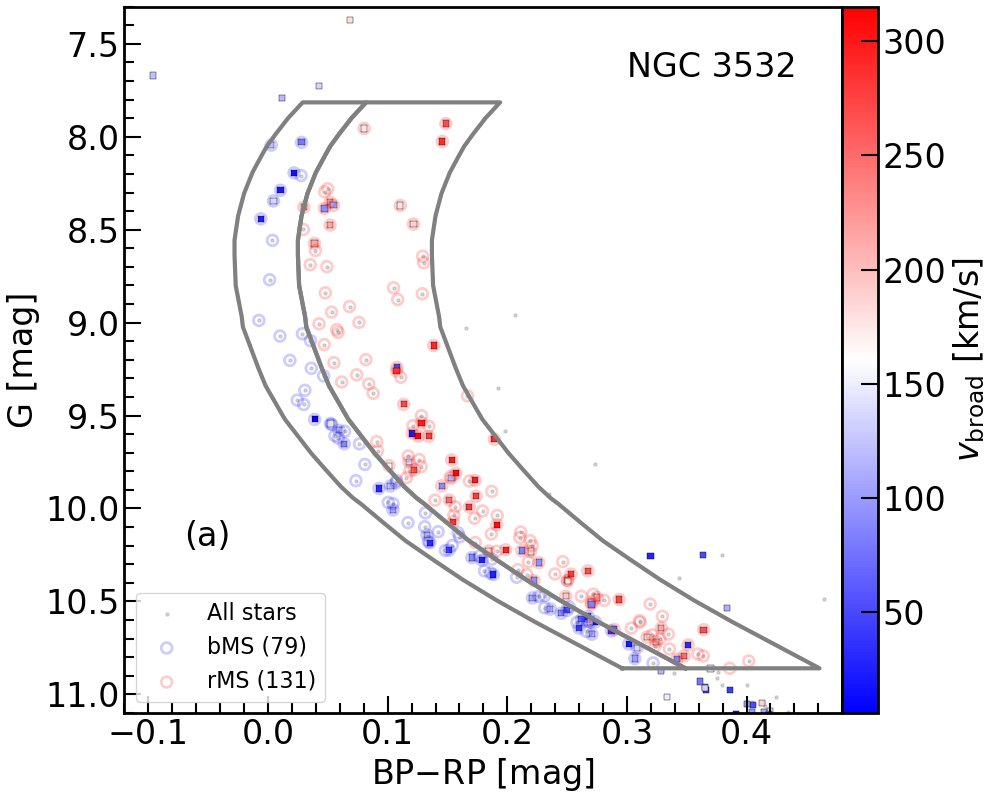}
    \includegraphics[width=0.98\linewidth]{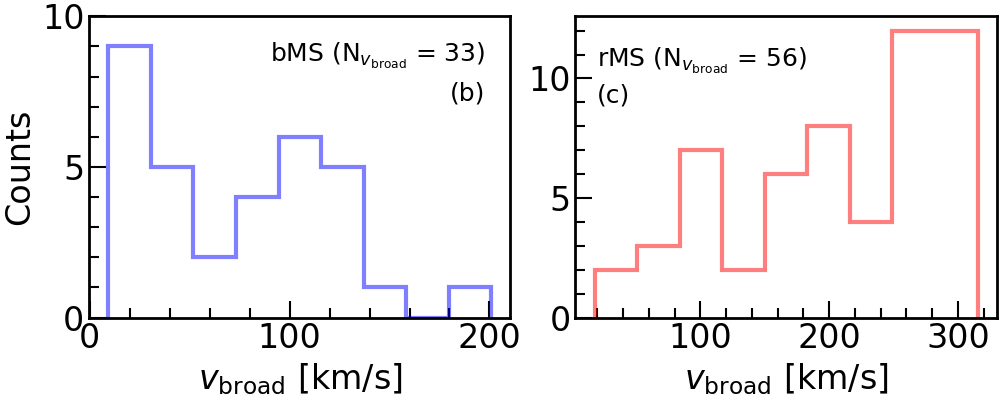}
    \caption{(a) The upper MS of the NGC\,3532 open cluster with the colourbar representing the $v_{\rm broad}$ measurements from the Gaia DR3 data. The filled coloured squares show members with available $v_{\rm broad}$ measurements. The selected bMS and rMS members are shown as blue and red open circles, respectively. (b) and (c) represent $v_{\rm broad}$ distributions of bMS and rMS members, respectively. }
    \label{fig:bMS_rMS}
\end{center}
\end{figure}

%tab01
\begin{table*}
\centering
\caption{Initial conditions of N-body simulations that model NGC 3532. The quantities $\mcl(0)$, $W_0(0)$, $\rh(0)$ and $\fbin(0)$ are the initial total cluster mass, King concentration parameter, half-mass radius, and primordial binary fraction, respectively. The abbreviation DM91 denotes the \citet{Duq_1991} orbital period distribution.}
\label{tab:NGC3532_models}
\begin{tabular}{cccccc}
\hline
Model no. & $\mcl(0)$ [$\Ms$] & $W_0(0)$ & $\rh(0)$ [pc] & $\fbin(0)$ & prim.\ binary distribution \\
\hline
1 & 4600 & 7 & 5.0 & 0.32 & DM91 \\
2 & 4600 & 7 & 5.0 & 0.32 & DM91 \\
3 & 5000 & 7 & 6.3 & 0.32 & DM91 \\
4 & 5000 & 7 & 6.3 & 0.32 & DM91 \\
5 & 5000 & 7 & 6.3 & 0.28 & DM91 \\
6 & 5000 & 7 & 6.3 & 0.28 & DM91 \\
7 & 5000 & 7 & 6.3 & 0.30 & DM91 \\
8 & 5000 & 7 & 6.3 & 0.30 & DM91 \\
9 & 5000 & 7 & 6.3 & 0.30 & Flat [$0.5$ -- $1.0 \times 10^{3}$ AU] \\
10 & 5000 & 7 & 6.3 & 0.30 & Flat [$0.5$ -- $1.0 \times 10^{3}$ AU] \\
11 & 5000 & 7 & 6.3 & 0.30 & Flat [$0.5$ -- $1.0 \times 10^{3}$ AU] \\
12 & 5000 & 7 & 6.0 & 0.27 & Flat [$0.1$ -- $2.1 \times 10^{2}$ AU] \\
13 & 5000 & 7 & 6.0 & 0.27 & Flat [$0.1$ -- $2.1 \times 10^{2}$ AU] \\
14 & 5000 & 7 & 6.0 & 0.27 & Flat [$0.1$ -- $2.1 \times 10^{2}$ AU] \\
\hline
\end{tabular}
\end{table*}

We adopt the cluster members from \citet{Rao2026ApJ..1002..103R}, identified using the ML-MOC algorithm \citep{Agarwal2021} on \textit{Gaia} DR3 data \citep{GaiaDR32023A&A...674A...1G}. The ML-MOC is a machine learning approach that integrates the k-Nearest Neighbor (kNN) and Gaussian Mixture Model (GMM) algorithms, leveraging proper motions and parallaxes to determine cluster membership. A total of 2,172 members are identified within the cluster radius of 120 arcmin. The cluster radius corresponds to the distance from the cluster centre (RA = 166.396804~deg; Dec = $-58.712294$~deg), where the cluster member distribution becomes indistinguishable from the surrounding field population. Figure~\ref{fig:obs_CMD} shows the cluster CMD fitted with non-rotating PARSEC isochrone \citep{Nguyen2022A&A...665A.126N} of $\log{\rm (Age/yr)} = 8.52$, distance of 490~pc, metallicity of [Fe/H] = 0.0~dex, and extinction of $A_{\rm v} = 0.07$~mag \citep{Rao2026ApJ..1002..103R} is shown in Fig.~\ref{fig:obs_CMD}. The equal-mass binary isochrone obtained by shifting the fitted isochrone by $ G = -0.75$~mag closely reproduces the observed equal-mass binary sequence along the MS in the cluster CMD. As can be seen from the inset figure (Figure~\ref{fig:obs_CMD}), the cluster clearly exhibits eMS. 

To estimate the total fraction of bMS stars in NGC\,3532, we identify the blue main-sequence (bMS) and red main-sequence (rMS) populations using the fitted isochrone together with the observed $v_{\rm broad}$ measurements of MS stars as shown in Fig.~\ref{fig:bMS_rMS}a. \citet{Cordoni2024MNRAS.532.1547C} demonstrated that $v_{\rm broad}$ measurements from Gaia DR3 can be reliably used as proxies for $v \sin i$ measurements based on comparison with the literature. Therefore, in this work we use $v_{\rm broad}$ measurements derived from Gaia DR3 spectra as the proxies for $v \sin i$. We define the boundary between the bMS and rMS by shifting the fitted single stars isochrone by 0.37 mag in colour. The faint limit of $G=11$~mag is chosen where the separation between slow and fast rotators remains clearly visible, while the bright limit corresponds to the MS turnoff. The majority of the slow and fast rotators are located within the selected bMS and rMS regions, respectively. A small number of slow rotators lie within the rMS region, which we discuss further in \S~\ref{sec:discussion}. We identify 79 bMS and 131 rMS stars, implying that the bMS accounts for approximately 37\% of the intermediate-mass MS population. The $v_{\rm broad}$ distribution of bMS and rMS is shown in Figs.~\ref{fig:bMS_rMS}(b) and \ref{fig:bMS_rMS}(c). Although $v_{\rm broad}$ measurements are available for only approximately 42\% of the stars in each population, the $v_{\rm broad}$ distributions of bMS and rMS clearly differ. A two-sample Anderson-Darling test confirms that the two distributions differ significantly (AD statistic = 5.13, $p=0.003$), with mean $v_{\rm broad}$ values of 76~km~s$^{-1}$ and 205~km~s$^{-1}$ for the bMS and rMS populations, respectively.

Having established the observed properties of the intermediate-mass cluster members, we next derive the present-day structural parameters and mass required to initialize the N-body simulations. We determine its core radius ($r_c$), half-mass radius ($r_h$), tidal radius ($r_t$), and the present-day mass of the cluster within the tidal radius ($M_{\rm cl}$) by fitting a single-mass isotropic King model \citet{King_1966} to the observed number density profile of the cluster following the procedure described by\citet{Rao2023MNRAS.526.1057R} (see Appendix~\ref{sec:structural_params}). The estimated parameters of the cluster are as follows: $r_{\rm c} = 15.62 \pm 0.66$~arcmin, $\rh = 42.33 \pm 4.38$~arcmin, $r_{\rm t} = 292.94 \pm 53.74$~arcmin, $\mcl = 2545 \pm 218$~M$_\odot$. At a distance of 490~pc, this corresponds to a projected half-mass radius ($\rh$) of $6.03 \pm 0.62$~pc. Based on different memberships, the binary fraction ($\fbin$) of NGC\,3532 varies from 0.16--0.27 in literature\citep{Clem2011AJ....141..115C, Cordoni2023A&A...672A..29C}, with a high-mass-ratio binary fraction as 0.22 \citep{Rao2026ApJ..1002..103R}. 

\section{N-body modelling of NGC\,3532}\label{nbody}

To test whether NGC\,3532 harbour a significant number of stellar merger products, which can lead to the formation of bMS of the cluster, we compute evolutionary star cluster models to reproduce the cluster's inferred present-day mass ($\mcl = 2545 \pm 218$~M$_\odot$), size ($\rh \approx 6.03 \pm 0.62$~pc), and $\fbin$ (0.22 -- 0.27). 
The star cluster models are computed with the direct N-body approach 
\citep{Aarseth_2003}. Below, we describe the computed models and the N-body code, thereafter the outcomes.

By trial and error, we probe a range of initial cluster models so that the key relevant observable parameters of NGC\,3532,
namely, its total photometric mass, size, and binary fraction, can be reasonably
(\ie, within either the measurement uncertainties or 10\%) reproduced at its inferred age of $\approx10^{8.52}$ years.
In this study, we limit ourselves to monolithic initial profiles, since NGC\,3532 has, overall, a smooth profile, despite
being highly extended (see Appendix.~\ref{sec:structural_params}). We choose King initial profiles \citep{King_1966} with the dimensionless concentration
parameter $W_0=7$. The initial profiles of our best-fitted models are listed in Table.~\ref{tab:NGC3532_models}.
We stress that these initial models do not comprise an exhaustive set of initial conditions that reproduce the properties of NGC\,3532.

The initial models comprise a primordial binary fraction of $\approx30$\% (Table.~\ref{tab:NGC3532_models}).
Two types of primordial binary distributions are considered. In one set of models, the widely used binary period distribution of
\citet[][hereafter DM91]{Duq_1991} is adopted. In another set of models, a flat semi-major-axis (hereafter SMA) distribution over
a wide range is adopted. The motivation for exploring different binary distributions is that the cluster's binary population
plays a key role in triggering star-star mergers by forming triple systems via binary-binary interactions \citep{Heggie_2003, Banerjee_2018}. Such
triples would undergo von Zeipel-Kozai-Lidov (hereafter ZKL; \citealt{vonZeipel_1910,Kozai_1962,Lidov_1962})
oscillations and/or chaotic internal evolution \citep{Trani_2024} and cause their member stars to merge.
In all cases, the member masses of the primordial binaries are paired by choosing masses from the stellar initial mass function
(see below). In all cluster models, the initial binary population follows a thermal eccentricity distribution \citep{Spitzer_1987}.

The clusters' initial population of member stars comprises zero-age-main-sequence (ZAMS) stars between $0.08-150\Ms$
sampled from the canonical initial mass function \citep[][hereafter IMF]{Kroupa_2001}. The clusters' stars are assumed to be homogeneous
in metallicity (young massive and open clusters do not exhibit a significant metallicity spread; \citealt{PortegiesZwart_2010})
with $Z$-metallicity of $\Zs$. The initial models are taken to be unsegregated.
All computed model clusters are subjected to a solar-neighbourhood-like external galactic field \citep{Allen_1991} during their evolution.

\subsection{Direct N-body evolution of model star clusters}\label{nbevol}

The model clusters are evolved with the state-of-the-art direct N-body evolution code $\nbseven$ \citep{Aarseth_2012}. {\tt NBODY7} is a variant of {\tt NBODY6} \citep{Nitadori_2012} that utilises {\tt OpenMP}-parallel graphical processing units (GPUs) to accelerate the regular force calculations and {\tt OpenMP}-parallel CPU threads to compute the irregular forces. The singularity due to the diverging gravitational force between the members of a hard binary or of a close hyperbolic encounter is dealt with by applying the Kustaanheimo-Stiefel regularization (KS-regularization). We refer to \cite{Aarseth_2003,Aarseth_2012} for details of the algorithms and numerical methods applied in {\tt NBODY6} and {\tt NBODY7}. The key difference between {\tt NBODY6} and {\tt NBODY7} is that the latter applies the Algorithmic Regularisation Chain ({\tt ARCHAIN}) \citep{Mikkola_1999} instead of the classical {\tt KS-Chain} \citep{Mikkola_1993}, to regularize triple and higher-order hierarchical systems inside the cluster. For compact hierarchical systems comprising high member mass ratios, which can frequently form via close dynamical interactions in high-binary-fraction clusters like those modelled here, {\tt ARCHAIN} has proven to be generally more accurate and stable.

The {\tt NBODY7} code incorporates stellar and binary evolution by interfacing with the semi-analytical binary evolution code {\tt BSE} \citep{Hurley_2002}. This allows taking into account the effects of evolutionary mass loss and remnant formation for every single- and binary-star member of the cluster. In the present computations, the \citet{Vink_2001} stellar wind model, black hole (BH) and neutron star (NS) formation via the `rapid' remnant mass and fallback models of \citet{Fryer_2012}, and the momentum-conserving remnant natal kick model are adopted. See \citet{Banerjee_2020,Banerjee_2020c} for detailed descriptions of these updates to {\tt NBODY7}. Due to the low escape speed of the present models, only a few BHs and NSs are retained inside them at formation, and they did not produce any general-relativistic mergers.

As for stellar collisions, the original {\tt BSE} recipes for collision remnants, as described in \citet[][see also \citealt{Hurley_2005}]{Hurley_2002} are preserved in {\tt NBODY7}. The main approximation here is the composition of the collision product's structure and the assignment of its rejuvenated age by assuming complete mixing.
In the present computations, 20\% of the secondary's mass
is assumed to be lost in a star-star collision (see \citealt{Banerjee_2020}), based on hydrodynamical studies of stellar collision
\citep{Lombardi_2002,Gaburov_2008,Glebbeek_2009,DeMink_2009}. 
All model clusters are evolved up to $\approx500$ Myr, well past the estimated age of $\approx330$ Myr of NGC\,3532.

\subsection{Results}\label{nbres}
%
%fig03
\begin{figure}
\begin{center}
\includegraphics[width=0.95\linewidth]{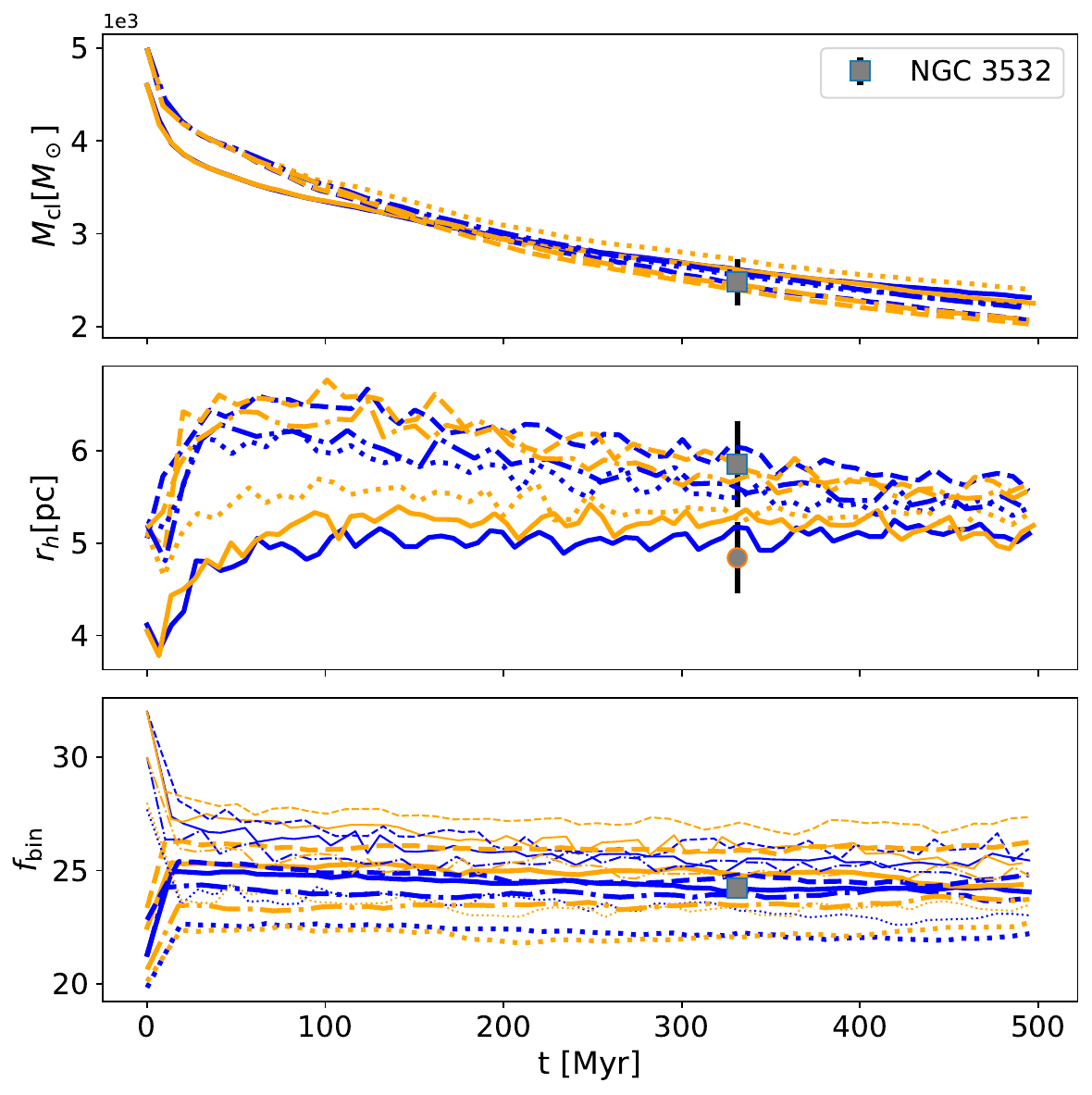}
\caption{Time evolution of model star clusters that reproduce the present-day mass, size (projected half-mass radius, averaged over three orthogonal projections), and binary fraction of the open star cluster NGC 3532. On each panel, the corresponding observed quantity is plotted (filled symbol). In the $\rh$ panel, the two symbols correspond to the two estimated distances of the cluster, namely, 405 pc and 490 pc. In the $\fbin$ panel, the thin (thick) lines correspond to all (hard/regularised) binaries inside the cluster at a given evolutionary time.}
\label{fig:Nbevol1}
\end{center}
\end{figure}

Fig.~\ref{fig:Nbevol1} shows the time evolution of the model clusters with the DM91 primordial binary distribution (Sec.~\ref{nbevol}, Table.~\ref{tab:NGC3532_models}). Shown are the time evolutions of the total cluster mass, $\mcl$, $\rh$,
(averaged over three orthogonal projections), and $\fbin$.
The observationally determined parameters of NGC\,3532 are over-plotted. The models show
good agreement with the observed parameters; at the age of the cluster, all models lie within either observational uncertainties or 
10\% (in the case of $\fbin$). Fig.~\ref{fig:Nbevol2} demonstrates the same for the model clusters initiating with the
flat primordial-binary distributions (Table.~\ref{tab:NGC3532_models}).

%fig04
\begin{figure*}
\centering
\includegraphics[width=0.95\linewidth]{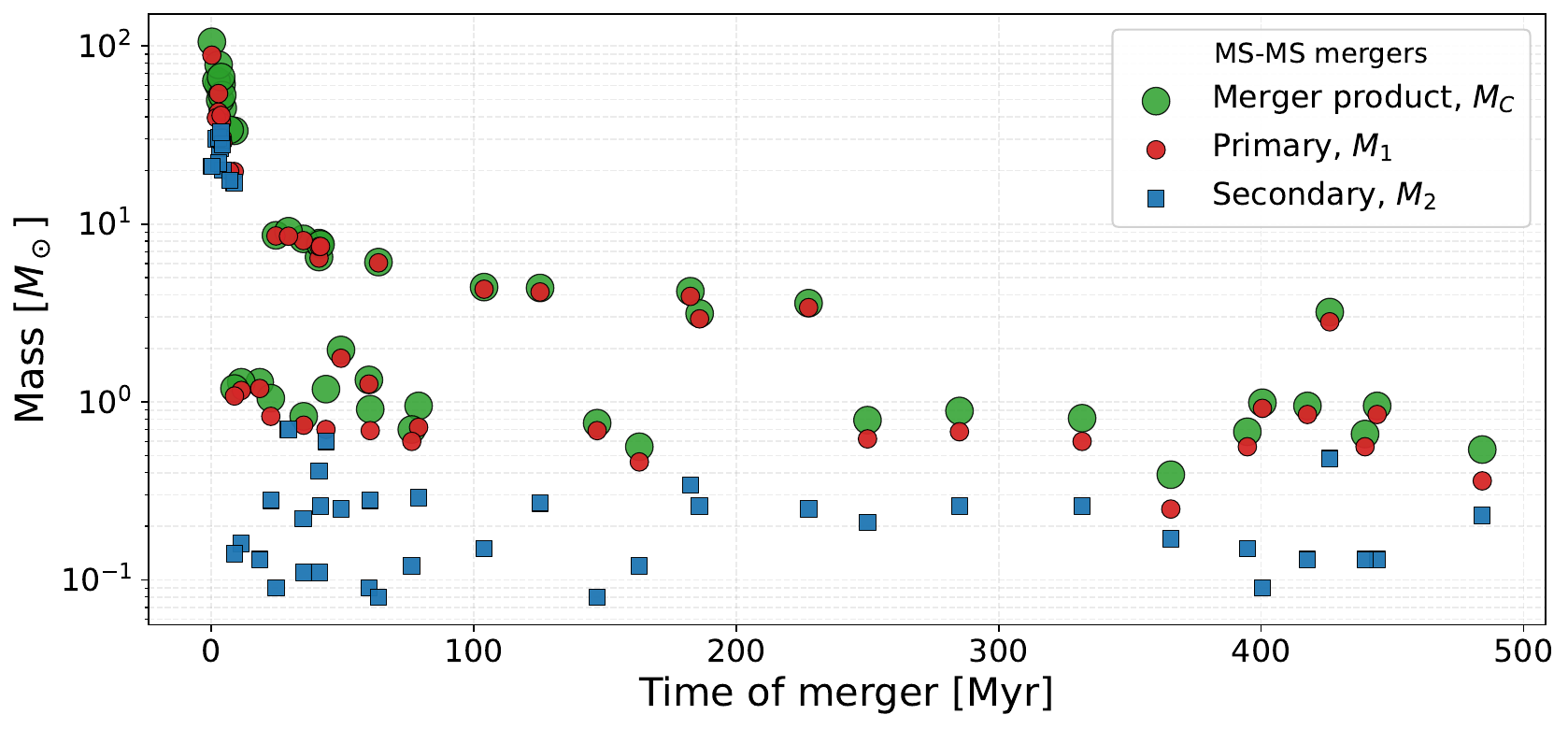}
\caption{Main-sequence--main-sequence mergers from all the computed N-body models. Shown in the Y-axis are the primary
mass, $\mone$, secondary mass, $\mtwo$ ($\mone>\mtwo$), and mass of the merger product, $\mc$ (legend),
for each merger event. The plotted masses correspond to the time of the respective merger event since the beginning of
the parent cluster's evolution, which are plotted along the X-axis.}
\label{fig:mergers}
\end{figure*}

Star-star mergers are of particular interest in this study, as stellar merger products, particularly MS-MS merger products, could be member candidates of
the slow-rotating bMS in NGC\,3532 (Sec.~\ref{sec:intro}). The $\nbseven$ code used in this work (Sec.~\ref{nbevol}) tracks and records all types of
collision and merger events runtime. Fig.~\ref{fig:mergers} shows all recorded MS-MS mergers from all the computed NGC\,3532 models
in Table.~\ref{tab:NGC3532_models}. In the following, we shall mainly restrict to only MS-MS mergers; so, for brevity, the term "merger" will imply an MS-MS merger unless stated otherwise.

The mergers happen in three main pathways, namely, (a) in-orbit collision due to a binary's eccentricity gain because of experiencing
dynamical interactions and/or ZKL oscillation and/or chaotic interaction inside a triple or higher-order compact subsystem,
(b) mass transfer in a binary due to the component stars' evolution leading to Roche lobe overflow (hereafter RLO),
and (c) RLO of the innermost binary driven by ZKL or chaotic interactions
\footnote{As arranged in $\nbseven$, these merger channels are separately recoded in the output files {\tt COLL}, {\tt COAL}, and {\tt COAL2}, respectively.}.
In the present models, all recorded mergers are mergers between the two members of a primordial binary.

Table~\ref{tab:NGC3532_mrg_eff} lists the number of mergers, $\nmrg$, and the corresponding merger efficiency, $\etamrg$, for each
of the NGC\,3532 models. Here, $\etamrg \equiv \nmrg/\mcl(0)$. Note that Fig.~\ref{fig:mergers} and Table~\ref{tab:NGC3532_mrg_eff} exclude
any merger declared right at $t=0$ evolutionary time, which is due to the high initial eccentricity of a small fraction of primordial binaries and is, therefore, an 
artefact of the specific instance of the initial condition. As seen in these illustrations, each computed model produces only a handful of mergers until
the $\approx330$ Myr age of NGC\,3532. In fact, the models with the flat initial SMA distributions, which contain much wider binaries
than the initially DM91 models (where the SMA peaks at $\sim 30$ AU), produce very few to no mergers. While wider binaries project a larger
geometrical cross section for close binary-single and binary-binary encounters, they also tend to ionise or need more extreme eccentricity
to undergo a collision among members, effectively inhibiting mergers. Overall, in order to match the moderate mass and extended size of NGC\,3532, all
the computed models have poor density, resulting in generally low dynamical interaction and merger rates.

%tab02
\begin{table}
\centering
\caption{Main-sequence--main-sequence (MS-MS) mergers of the computed NGC 3532 models in Table~\ref{tab:NGC3532_models}.
The quantity $\nmrg$ is the total number of MS-MS mergers in the model and $\etamrg$ is the corresponding merger efficiency (see text).}
\label{tab:NGC3532_mrg_eff}
\begin{tabular}{ccc}
\hline
Model no. & $\nmrg$ & $\etamrg$ [M$_\odot^{-1}$] \\
\hline
1 & 5 & $1.1 \times 10^{-3}$ \\
2 & 8 & $1.7 \times 10^{-3}$ \\
3 & 5 & $1.0 \times 10^{-3}$ \\
4 & 5 & $1.0 \times 10^{-3}$ \\
5 & 6 & $1.2 \times 10^{-3}$ \\
6 & 4 & $8.0 \times 10^{-4}$ \\
7 & 6 & $1.2 \times 10^{-3}$ \\
8 & 6 & $1.2 \times 10^{-3}$ \\
9 & 1 & $2.0 \times 10^{-4}$ \\
10 & 0 & 0 \\
11 & 0 & 0 \\
12 & 0 & 0 \\
13 & 0 & 0 \\
14 & 1 & $2.0 \times 10^{-4}$ \\
\hline
\end{tabular}
\end{table}

%-------------- Discussion ------------------%

\section{Discussion} \label{sec:discussion}

In this study, we use N-body simulations to investigate the frequency of stellar merger products and assess their possible contribution to the formation of bMS stars in the open cluster NGC\,3532. This cluster contains approximately 37\% bMS stars among its eMS population, which is predominantly composed of slow rotators. Our simulations demonstrate that the production rate of merger products is very low, even when considering all 14 models. Most merger products have either already evolved away from the main-sequence region, given the cluster turnoff mass of approximately 3.1~M$_\odot$, or are located among the low-mass cluster population ($<1.2$~M$_\odot$). Consequently, only a small number of merger products remain within the current intermediate-mass range (1.5--3.1~M$_\odot$), corresponding to the eMS region of NGC~3532. Therefore, the number of merger products produced in our simulations is insufficient to explain the observed fraction of bMS stars.

Following a merger event, stellar merger products are generally expected to become bluer and more luminous due to the replenishment of hydrogen in their cores. They are also expected to rotate rapidly as a consequence of angular momentum conservation during the merger process. Interestingly, all merger events producing intermediate-mass merger products in our simulations occur after 100~Myr. This implies that additional mechanisms would be required to remove a substantial amount of angular momentum from these stars within only 100--200~Myr in order to transform them into the observed slow-rotating bMS population.

Our simulations indicate that NGC\,3532 undergoes rapid mass loss, losing approximately half of its initial mass within $\sim330$~Myr.
Following the initial expansion caused by wind and supernova mass loss from massive stars, however, the cluster radius changes only marginally, resulting in a persistently low-density environment. The reduced stellar density limits dynamical interactions between binary systems and between binaries and single stars, thereby reducing the probability of merger events. Consequently, the low frequency of merger products in our simulations cannot account for the observed bMS population. Nevertheless, the efficiency of producing merger products is expected to vary significantly among clusters depending on their initial binary fraction, stellar density, and dynamical evolution.

It is also important to consider that merger events occur at different evolutionary stages, and the merger products may acquire different amounts of mass, resulting in a range of positions in the CMD. However, NGC~3532 exhibits a remarkably narrow and well-defined MS and hosts only one blue straggler star located significantly above and bluer than the MS turnoff. This blue straggler is a slow rotator based on its $v_{\rm broad}$ measurement (Fig.~\ref{fig:bMS_rMS}a), suggesting that it may have formed relatively early in the cluster evolution and subsequently experienced significant angular momentum loss.

Recently, \citep{Dvo2024A&A...689A.234D} performed N-body simulations of clusters with initial masses ranging from 67.5~M$\odot$ to 8000~M$\odot$ to estimate the contribution of stellar mergers for different spectral types. Their simulations assumed an initial binary fraction of 100\% for all cluster members. They found that approximately 12.54--23.24\% of B-type stars and 4.8--7.7\% of A-type stars could originate from stellar mergers, populations that typically overlap with the eMS and eMSTO regions. However, even under the extreme assumption of a fully binary cluster, their simulations could not reproduce the typical observed fraction of bMS stars. This further suggests that stellar mergers alone are unlikely to be the dominant mechanism responsible for producing bMS populations in open clusters with similar densities.

The slow rotation observed in NGC~3532 likely requires mechanisms capable of removing angular momentum during the pre-main-sequence phase, in addition to tidal synchronization in binary systems. As shown in Figure~\ref{fig:bMS_rMS}(a), the majority of slow rotators are located along the bMS, although a small number of slow rotators are also present in the rMS region. Since NGC~3532 suffers from negligible foreground extinction, the observed photometric positions of both slow and fast rotators likely represent their intrinsic locations in the CMD.

The slow rotators located in the rMS region may have a binary-related origin, as this region is typically associated with high mass-ratio binaries. NGC\,3532 contains seven known eclipsing and spectroscopic binaries \citep{Rao2026ApJ..1002..103R}. Although rotational velocities are unavailable for all systems, three are confirmed slow rotators. Furthermore, two of these binaries have high mass ratios and orbital periods shorter than 10 days, consistent with the expected timescale for tidal synchronization within the cluster lifetime. In addition, NGC~3532 hosts 11 variables classified as ACV~$\mid$~CP~$\mid$~MCP~$\mid$~ROAM~$\mid$~ROAP~$\mid$~SXARI variables \citep{Rao2025arXiv251205458R}. These objects exhibit photometric variability commonly associated with surface abundance spots. Chemically peculiar Ap stars are known to possess strong surface magnetic fields and typically rotate slowly. Although magnetic fields have not yet been directly confirmed for these 11 stars, their formation may be related to fossil magnetic fields \citep{Braithwaite2004Natur.431..819B} or merger processes \citep{Bogomazov2009ARep...53..214B, Schneider2016MNRAS.457.2355S, Hubrig2025A&A...701A.255H}. These variables are distributed across both the bMS and rMS regions. However, those located in the rMS generally have RUWE values greater than 1.2, suggesting possible unresolved companions. Therefore, slow rotators in the rMS may plausibly result from tidal interactions in binary systems. In contrast, the slow rotators located in the bMS region likely require alternative formation pathways, such as early angular momentum loss during the pre-main-sequence phase, magnetic braking associated with fossil magnetic fields, or tidal effects in low mass-ratio binaries.

%-------------- Conclusion & Summary ------------------%

\section{Summary}\label{sec:summary}
The different rotational properties of intermediate-mass stars are now recognized as the primary origin of the observed eMS and eMSTO features in young and intermediate-age clusters. While fast rotators can be naturally explained by the evolution of stars with radiative envelopes, the origin of slowly rotating populations remains an open question. \citet{Wang2022NatAs...6..480W} proposed that stellar merger products formed during early cluster evolution can subsequently lose angular momentum and evolve into the bMS population. However, several observational and theoretical challenges remain for this scenario \citep{Bastian2025A&A...700A.241B, Schneider2025arXiv250918421S}

In this study, we performed a set of evolutionary simulations of model star clusters using the $\nbseven$ direct N-body code to reproduce several key present-day properties of the well-studied open cluster NGC\,3532. The cluster has an age of approximately 330~Myr and lies along a line of sight with very low foreground extinction, making it an ideal laboratory for studying the intrinsic relationship between stellar rotation and CMD morphology. Our simulations show that only a small fraction of merger products remain within the intermediate-mass main-sequence region at the present cluster age. The observed bMS population accounts for approximately 37\% of the eMS stars, whereas the number of merger products produced in our simulations is far too small to explain this fraction. This suggests that stellar mergers are unlikely to be the dominant mechanism responsible for producing the bMS population in NGC\,3532. Since open clusters are generally low-density environments with relatively large sizes, their dynamical evolution and binary populations may not produce sufficient numbers of mergers to explain the observed slow-rotating bMS population. Instead, mechanisms capable of removing angular momentum at early evolutionary stages, particularly during the pre-main-sequence phase, are likely required to explain the origin of slow rotators among intermediate-mass stars in open clusters.
Future studies combining detailed magnetic field measurements, binary population analysis, and rotational velocities of star-forming clusters and young clusters containing intermediate-mass stars in the pre-main-sequence phase are essential to determine the dominant mechanism responsible for the formation of bMS populations.
%%%%%%%%%%%%%%%%%%%%%%%%%%%%%%%%%%%%%%%%%%%%%%%%%%

\begin{acknowledgements}
    KKR acknowledges funding from the National Science and Technology Council of Taiwan (NSTC~113-2123-M-008-004).     
    SB acknowledges funding for this work by the Deutsche Forschungsgemeinschaft (DFG; German Research Foundation) through the project ``The dynamics of stellar-mass black holes in
    dense stellar systems and their role in gravitational wave generation'' (project number 405620641; PI: S. Banerjee). The $N$-body simulations have been carried out on the {\tt gpudyn} GPU computing servers containing NVIDIA Ampere A40 and NVIDIA RTX 2080 GPUs, located at the Argelander-Institut f\"ur Astronomie (AIfA), University of Bonn.
    The {\tt gpudyn} servers have been sponsored by the above-mentioned DFG project, HISKP, and the University of Bonn. SB thanks the AIfA for
    hospitality.
    This work includes data from the ESO Science Archive Facility (\url{https://doi.org/10.18727/archive/25}) and the third data release from the European Space Agency (ESA) mission {\it Gaia} \citep[\url{https://www.cosmos.esa.int/gaia};][]{GaiaDR32023A&A...674A...1G}, processed by the {\it Gaia} Data Processing and Analysis Consortium (DPAC, \url{https://www.cosmos.esa.int/web/gaia/dpac/consortium}). We also used the Astrophysics Data System (ADS) governed by NASA (\url{https://ui.adsabs.harvard.edu}).
    The following tools are also used in this work: 
    \textsc{Astropy} \citep{2018AJ....156..123A}; 
    \textsc{Astroquery} \citep{Ginsburg2019AJ....157...98G};
    \textsc{Matplotlib} \citep{Hunter:2007};
    \textsc{Nbody7} \citep{Aarseth_2012};
    \textsc{NumPy} \citep{2020Natur.585..357Harris}.
\end{acknowledgements}

%%%%%%%%%%%%%%%%%%%%%%%%%%%%%%%%%%%%%%%%%%%%%%%%%%%%%%%%%%%%%%
\bibliographystyle{aa_url}
\bibliography{references}

@ARTICLE{2018AJ....156..123A,
       author = {{Astropy Collaboration} and {Price-Whelan}, A.~M. and {Sip{\H{o}}cz}, B.~M. and {G{\"u}nther}, H.~M. and {Lim}, P.~L. and {Crawford}, S.~M. and {Conseil}, S. and {Shupe}, D.~L. and {Craig}, M.~W. and {Dencheva}, N. and {Ginsburg}, A. and {VanderPlas}, J.~T. and {Bradley}, L.~D. and {P{\'e}rez-Su{\'a}rez}, D. and {de Val-Borro}, M. and {Aldcroft}, T.~L. and {Cruz}, K.~L. and {Robitaille}, T.~P. and {Tollerud}, E.~J. and {Ardelean}, C. and {Babej}, T. and {Bach}, Y.~P. and {Bachetti}, M. and {Bakanov}, A.~V. and {Bamford}, S.~P. and {Barentsen}, G. and {Barmby}, P. and {Baumbach}, A. and {Berry}, K.~L. and {Biscani}, F. and {Boquien}, M. and {Bostroem}, K.~A. and {Bouma}, L.~G. and {Brammer}, G.~B. and {Bray}, E.~M. and {Breytenbach}, H. and {Buddelmeijer}, H. and {Burke}, D.~J. and {Calderone}, G. and {Cano Rodr{\'\i}guez}, J.~L. and {Cara}, M. and {Cardoso}, J.~V.~M. and {Cheedella}, S. and {Copin}, Y. and {Corrales}, L. and {Crichton}, D. and {D'Avella}, D. and {Deil}, C. and {Depagne}, {\'E}. and {Dietrich}, J.~P. and {Donath}, A. and {Droettboom}, M. and {Earl}, N. and {Erben}, T. and {Fabbro}, S. and {Ferreira}, L.~A. and {Finethy}, T. and {Fox}, R.~T. and {Garrison}, L.~H. and {Gibbons}, S.~L.~J. and {Goldstein}, D.~A. and {Gommers}, R. and {Greco}, J.~P. and {Greenfield}, P. and {Groener}, A.~M. and {Grollier}, F. and {Hagen}, A. and {Hirst}, P. and {Homeier}, D. and {Horton}, A.~J. and {Hosseinzadeh}, G. and {Hu}, L. and {Hunkeler}, J.~S. and {Ivezi{\'c}}, {\v{Z}}. and {Jain}, A. and {Jenness}, T. and {Kanarek}, G. and {Kendrew}, S. and {Kern}, N.~S. and {Kerzendorf}, W.~E. and {Khvalko}, A. and {King}, J. and {Kirkby}, D. and {Kulkarni}, A.~M. and {Kumar}, A. and {Lee}, A. and {Lenz}, D. and {Littlefair}, S.~P. and {Ma}, Z. and {Macleod}, D.~M. and {Mastropietro}, M. and {McCully}, C. and {Montagnac}, S. and {Morris}, B.~M. and {Mueller}, M. and {Mumford}, S.~J. and {Muna}, D. and {Murphy}, N.~A. and {Nelson}, S. and {Nguyen}, G.~H. and {Ninan}, J.~P. and {N{\"o}the}, M. and {Ogaz}, S. and {Oh}, S. and {Parejko}, J.~K. and {Parley}, N. and {Pascual}, S. and {Patil}, R. and {Patil}, A.~A. and {Plunkett}, A.~L. and {Prochaska}, J.~X. and {Rastogi}, T. and {Reddy Janga}, V. and {Sabater}, J. and {Sakurikar}, P. and {Seifert}, M. and {Sherbert}, L.~E. and {Sherwood-Taylor}, H. and {Shih}, A.~Y. and {Sick}, J. and {Silbiger}, M.~T. and {Singanamalla}, S. and {Singer}, L.~P. and {Sladen}, P.~H. and {Sooley}, K.~A. and {Sornarajah}, S. and {Streicher}, O. and {Teuben}, P. and {Thomas}, S.~W. and {Tremblay}, G.~R. and {Turner}, J.~E.~H. and {Terr{\'o}n}, V. and {van Kerkwijk}, M.~H. and {de la Vega}, A. and {Watkins}, L.~L. and {Weaver}, B.~A. and {Whitmore}, J.~B. and {Woillez}, J. and {Zabalza}, V. and {Astropy Contributors}},
        title = "{The Astropy Project: Building an Open-science Project and Status of the v2.0 Core Package}",
      journal = {\aj},
         year = 2018,
        month = sep,
       volume = {156},
       number = {3},
          eid = {123},
        pages = {123},
          doi = {10.3847/1538-3881/aabc4f},
archivePrefix = {arXiv},
       eprint = {1801.02634},
 primaryClass = {astro-ph.IM},
       adsurl = {https://ui.adsabs.harvard.edu/abs/2018AJ....156..123A}
}

@ARTICLE{Agarwal2021,
       author = {{Agarwal}, Manan and {Rao}, Khushboo K. and {Vaidya}, Kaushar and {Bhattacharya}, Souradeep},
        title = "{ML-MOC: Machine Learning (kNN and GMM) based Membership determination for Open Clusters}",
      journal = {\mnras},
         year = 2021,
        month = apr,
       volume = {502},
       number = {2},
        pages = {2582-2599},
          doi = {10.1093/mnras/stab118},
archivePrefix = {arXiv},
       eprint = {2011.13622},
 primaryClass = {astro-ph.IM},
       adsurl = {https://ui.adsabs.harvard.edu/abs/2021MNRAS.502.2582A}
}

@ARTICLE{Dvo2024A&A...689A.234D,
       author = {{Dvo{\v{r}}{\'a}kov{\'a}}, N. and {Kor{\v{c}}{\'a}kov{\'a}}, D. and {Dinnbier}, F. and {Kroupa}, P.},
        title = "{The mass distribution of stellar mergers: A new scenario for several FS CMa stars}",
      journal = {\aap},
         year = 2024,
        month = sep,
       volume = {689},
          eid = {A234},
        pages = {A234},
          doi = {10.1051/0004-6361/202449586},
archivePrefix = {arXiv},
       eprint = {2410.01882},
 primaryClass = {astro-ph.GA},
       adsurl = {https://ui.adsabs.harvard.edu/abs/2024A&A...689A.234D}
}

@ARTICLE{Schneider2025arXiv250918421S,
       author = {{Schneider}, Fabian R.~N.},
        title = "{Theory, Simulations and Observations of Stellar Mergers}",
      journal = {arXiv e-prints},
         year = 2025,
        month = sep,
          eid = {arXiv:2509.18421},
        pages = {arXiv:2509.18421},
          doi = {10.48550/arXiv.2509.18421},
archivePrefix = {arXiv},
       eprint = {2509.18421},
 primaryClass = {astro-ph.SR},
       adsurl = {https://ui.adsabs.harvard.edu/abs/2025arXiv250918421S}
}

@ARTICLE{Wang2020ApJ...888L..12W,
       author = {{Wang}, Chen and {Langer}, Norbert and {Schootemeijer}, Abel and {Castro}, Norberto and {Adscheid}, Sylvia and {Marchant}, Pablo and {Hastings}, Ben},
        title = "{Effects of Close Binary Evolution on the Main-sequence Morphology of Young Star Clusters}",
      journal = {\apjl},
         year = 2020,
        month = jan,
       volume = {888},
       number = {1},
          eid = {L12},
        pages = {L12},
          doi = {10.3847/2041-8213/ab6171},
archivePrefix = {arXiv},
       eprint = {1912.07294},
 primaryClass = {astro-ph.SR},
       adsurl = {https://ui.adsabs.harvard.edu/abs/2020ApJ...888L..12W}
}

@ARTICLE{Cordoni2024MNRAS.532.1547C,
       author = {{Cordoni}, G. and {Casagrande}, L. and {Yu}, J. and {Milone}, A.~P. and {Marino}, A.~F. and {D'Antona}, F. and {Dell'Agli}, F. and {Buder}, S. and {Tailo}, M.},
        title = "{Survey of extended main-sequence turn-offs in galactic open clusters: stellar rotations from Gaia RVS spectra}",
      journal = {\mnras},
         year = 2024,
        month = aug,
       volume = {532},
       number = {2},
        pages = {1547-1563},
          doi = {10.1093/mnras/stae1569},
archivePrefix = {arXiv},
       eprint = {2406.16551},
 primaryClass = {astro-ph.GA},
       adsurl = {https://ui.adsabs.harvard.edu/abs/2024MNRAS.532.1547C}
}

@ARTICLE{Deng2024RAA....24f5004D,
       author = {{Deng}, Yang-Yang and {Li}, Zhong-Mu},
        title = "{Study of 26 Galactic Open Clusters with Extended Main-sequence Turnoffs}",
      journal = {Research in Astronomy and Astrophysics},
         year = 2024,
        month = jun,
       volume = {24},
       number = {6},
          eid = {065004},
        pages = {065004},
          doi = {10.1088/1674-4527/ad3dc5},
archivePrefix = {arXiv},
       eprint = {2403.18234},
 primaryClass = {astro-ph.SR},
       adsurl = {https://ui.adsabs.harvard.edu/abs/2024RAA....24f5004D}
}

@ARTICLE{Marino2018ApJ...863L..33M,
       author = {{Marino}, A.~F. and {Milone}, A.~P. and {Casagrande}, L. and {Przybilla}, N. and {Balaguer-N{\'u}{\~n}ez}, L. and {Di Criscienzo}, M. and {Serenelli}, A. and {Vilardell}, F.},
        title = "{Discovery of Extended Main Sequence Turnoffs in Galactic Open Clusters}",
      journal = {\apjl},
         year = 2018,
        month = aug,
       volume = {863},
       number = {2},
          eid = {L33},
        pages = {L33},
          doi = {10.3847/2041-8213/aad868},
archivePrefix = {arXiv},
       eprint = {1807.05888},
 primaryClass = {astro-ph.SR},
       adsurl = {https://ui.adsabs.harvard.edu/abs/2018ApJ...863L..33M}
}

@ARTICLE{Milone2009A&A...497..755M,
       author = {{Milone}, A.~P. and {Bedin}, L.~R. and {Piotto}, G. and {Anderson}, J.},
        title = "{Multiple stellar populations in Magellanic Cloud clusters. I. An ordinary feature for intermediate age globulars in the LMC?}",
      journal = {\aap},
         year = 2009,
        month = apr,
       volume = {497},
       number = {3},
        pages = {755-771},
          doi = {10.1051/0004-6361/200810870},
archivePrefix = {arXiv},
       eprint = {0810.2558},
 primaryClass = {astro-ph},
       adsurl = {https://ui.adsabs.harvard.edu/abs/2009A&A...497..755M}
}

@ARTICLE{Bastian2009MNRAS.398L..11B,
       author = {{Bastian}, N. and {de Mink}, S.~E.},
        title = "{The effect of stellar rotation on colour-magnitude diagrams: on the apparent presence of multiple populations in intermediate age stellar clusters}",
      journal = {\mnras},
         year = 2009,
        month = sep,
       volume = {398},
       number = {1},
        pages = {L11-L15},
          doi = {10.1111/j.1745-3933.2009.00696.x},
archivePrefix = {arXiv},
       eprint = {0906.1590},
 primaryClass = {astro-ph.GA},
       adsurl = {https://ui.adsabs.harvard.edu/abs/2009MNRAS.398L..11B}
}

@ARTICLE{Bastian2018MNRAS.480.3739B,
       author = {{Bastian}, N. and {Kamann}, S. and {Cabrera-Ziri}, I. and {Georgy}, C. and {Ekstr{\"o}m}, S. and {Charbonnel}, C. and {de Juan Ovelar}, M. and {Usher}, C.},
        title = "{Extended main sequence turnoffs in open clusters as seen by Gaia - I. NGC 2818 and the role of stellar rotation}",
      journal = {\mnras},
         year = 2018,
        month = nov,
       volume = {480},
       number = {3},
        pages = {3739-3746},
          doi = {10.1093/mnras/sty2100},
archivePrefix = {arXiv},
       eprint = {1807.10779},
 primaryClass = {astro-ph.SR},
       adsurl = {https://ui.adsabs.harvard.edu/abs/2018MNRAS.480.3739B}
}

@ARTICLE{Kamann2020MNRAS.492.2177K,
       author = {{Kamann}, S. and {Bastian}, N. and {Gossage}, S. and {Baade}, D. and {Cabrera-Ziri}, I. and {Da Costa}, G. and {de Mink}, S.~E. and {Georgy}, C. and {Giesers}, B. and {G{\"o}ttgens}, F. and {Hilker}, M. and {Husser}, T.-O. and {Lardo}, C. and {Larsen}, S.~S. and {Mackey}, D. and {Martocchia}, S. and {Mucciarelli}, A. and {Platais}, I. and {Roth}, M.~M. and {Salaris}, M. and {Usher}, C. and {Yong}, D.},
        title = "{How stellar rotation shapes the colour-magnitude diagram of the massive intermediate-age star cluster NGC 1846}",
      journal = {\mnras},
         year = 2020,
        month = feb,
       volume = {492},
       number = {2},
        pages = {2177-2192},
          doi = {10.1093/mnras/stz3583},
archivePrefix = {arXiv},
       eprint = {2001.01731},
 primaryClass = {astro-ph.SR},
       adsurl = {https://ui.adsabs.harvard.edu/abs/2020MNRAS.492.2177K}
}

@ARTICLE{Kamann2023MNRAS.518.1505K,
       author = {{Kamann}, S. and {Saracino}, S. and {Bastian}, N. and {Gossage}, S. and {Usher}, C. and {Baade}, D. and {Cabrera-Ziri}, I. and {de Mink}, S.~E. and {Ekstrom}, S. and {Georgy}, C. and {Hilker}, M. and {Larsen}, S.~S. and {Mackey}, D. and {Niederhofer}, F. and {Platais}, I. and {Yong}, D.},
        title = "{The effects of stellar rotation along the main sequence of the 100-Myr-old massive cluster NGC 1850}",
      journal = {\mnras},
         year = 2023,
        month = jan,
       volume = {518},
       number = {1},
        pages = {1505-1521},
          doi = {10.1093/mnras/stac3170},
archivePrefix = {arXiv},
       eprint = {2211.00693},
 primaryClass = {astro-ph.SR},
       adsurl = {https://ui.adsabs.harvard.edu/abs/2023MNRAS.518.1505K}
}

@ARTICLE{Rivinius2013A&ARv..21...69R,
       author = {{Rivinius}, Thomas and {Carciofi}, Alex C. and {Martayan}, Christophe},
        title = "{Classical Be stars. Rapidly rotating B stars with viscous Keplerian decretion disks}",
      journal = {\aapr},
         year = 2013,
        month = oct,
       volume = {21},
          eid = {69},
        pages = {69},
          doi = {10.1007/s00159-013-0069-0},
archivePrefix = {arXiv},
       eprint = {1310.3962},
 primaryClass = {astro-ph.SR},
       adsurl = {https://ui.adsabs.harvard.edu/abs/2013A&ARv..21...69R}
}

@ARTICLE{DAntona2015MNRAS4532637,
  author = {D’Antona, F. and Vesperini, E. and D’Ercole, A. and D’Antona, F. and Tailo, M.},
  title = {Effects of rotation on the morphology of the colour–magnitude diagrams of intermediate‐age clusters},
  journal = {MNRAS},
  year = {2015},
  volume = {453},
  number = {3},
  pages = {2637–2651},
  doi = {10.1093/mnras/stv1791},
  archivePrefix = {arXiv},
  eprint = {1504.02253},
  primaryClass = {astro-ph.SR},
  adsurl = {https://ui.adsabs.harvard.edu/abs/2015MNRAS.453.2637D}
}

@ARTICLE{Wang2022NatAs...6..480W,
       author = {{Wang}, Chen and {Langer}, Norbert and {Schootemeijer}, Abel and {Milone}, Antonino and {Hastings}, Ben and {Xu}, Xiao-Tian and {Bodensteiner}, Julia and {Sana}, Hugues and {Castro}, Norberto and {Lennon}, D.~J. and {Marchant}, Pablo and {de Koter}, A. and {de Mink}, Selma E.},
        title = "{Stellar mergers as the origin of the blue main-sequence band in young star clusters}",
      journal = {Nature Astronomy},
         year = 2022,
        month = feb,
       volume = {6},
        pages = {480-487},
          doi = {10.1038/s41550-021-01597-5},
archivePrefix = {arXiv},
       eprint = {2202.05552},
 primaryClass = {astro-ph.SR},
       adsurl = {https://ui.adsabs.harvard.edu/abs/2022NatAs...6..480W}
}

@ARTICLE{Bastian2020MNRAS.495.1978B,
       author = {{Bastian}, Nate and {Kamann}, Sebastian and {Amard}, Louis and {Charbonnel}, Corinne and {Haemmerl{\'e}}, Lionel and {Matt}, Sean P.},
        title = "{On the origin of the bimodal rotational velocity distribution in stellar clusters: rotation on the pre-main sequence}",
      journal = {\mnras},
         year = 2020,
        month = jun,
       volume = {495},
       number = {2},
        pages = {1978-1983},
          doi = {10.1093/mnras/staa1332},
archivePrefix = {arXiv},
       eprint = {2005.01779},
 primaryClass = {astro-ph.SR},
       adsurl = {https://ui.adsabs.harvard.edu/abs/2020MNRAS.495.1978B}
}

@ARTICLE{Braithwaite2004Natur.431..819B,
       author = {{Braithwaite}, Jonathan and {Spruit}, Hendrik C.},
        title = "{A fossil origin for the magnetic field in A stars and white dwarfs}",
      journal = {\nat},
         year = 2004,
        month = oct,
       volume = {431},
       number = {7010},
        pages = {819-821},
          doi = {10.1038/nature02934},
archivePrefix = {arXiv},
       eprint = {astro-ph/0502043},
 primaryClass = {astro-ph},
       adsurl = {https://ui.adsabs.harvard.edu/abs/2004Natur.431..819B}
}

@ARTICLE{Schneider2016MNRAS.457.2355S,
       author = {{Schneider}, F.~R.~N. and {Podsiadlowski}, Ph. and {Langer}, N. and {Castro}, N. and {Fossati}, L.},
        title = "{Rejuvenation of stellar mergers and the origin of magnetic fields in massive stars}",
      journal = {\mnras},
         year = 2016,
        month = apr,
       volume = {457},
       number = {3},
        pages = {2355-2365},
          doi = {10.1093/mnras/stw148},
archivePrefix = {arXiv},
       eprint = {1601.05084},
 primaryClass = {astro-ph.SR},
       adsurl = {https://ui.adsabs.harvard.edu/abs/2016MNRAS.457.2355S}
}

@ARTICLE{Hubrig2025A&A...701A.255H,
       author = {{Hubrig}, S. and {J{\"a}rvinen}, S.~P. and {Ilyin}, I. and {Sch{\"o}ller}, M.},
        title = "{The incidence of magnetism in blue and yellow straggler stars}",
      journal = {\aap},
         year = 2025,
        month = sep,
       volume = {701},
          eid = {A255},
        pages = {A255},
          doi = {10.1051/0004-6361/202556166},
archivePrefix = {arXiv},
       eprint = {2509.02304},
 primaryClass = {astro-ph.SR},
       adsurl = {https://ui.adsabs.harvard.edu/abs/2025A&A...701A.255H}
}

@ARTICLE{Bogomazov2009ARep...53..214B,
       author = {{Bogomazov}, A.~I. and {Tutukov}, A.~V.},
        title = "{Merging of components in close binaries: Type Ia supernovae, massive white dwarfs, and Ap stars}",
      journal = {Astronomy Reports},
         year = 2009,
        month = mar,
       volume = {53},
       number = {3},
        pages = {214-222},
          doi = {10.1134/S1063772909030032},
archivePrefix = {arXiv},
       eprint = {0901.4899},
 primaryClass = {astro-ph.SR},
       adsurl = {https://ui.adsabs.harvard.edu/abs/2009ARep...53..214B}
}

@ARTICLE{Duq_1991,
   author = {{Duquennoy}, A. and {Mayor}, M.},
    title = "{Multiplicity among solar-type stars in the solar neighbourhood. II - Distribution of the orbital elements in an unbiased sample}",
  journal = {\aap},
     year = 1991,
    month = aug,
   volume = 248,
    pages = {485-524},
   adsurl = {http://adsabs.harvard.edu/abs/1991A%26A...248..485D}
}

@BOOK{Aarseth_2003,
   author = {{Aarseth}, S.~J.},
    title = "{Gravitational N-Body Simulations, Cambridge University Press, Cambridge, UK, pp.~430.~ISBN 0521432723}",
booktitle = {Gravitational N-Body Simulations, by Sverre J.~Aarseth, pp.~430.~ISBN 0521432723.~Cambridge, UK: Cambridge University Press, November 2003.},
     year = 2003,
    month = oct,
    pages = {430},
   adsurl = {http://adsabs.harvard.edu/abs/2003gnbs.book.....A}
}

@ARTICLE{King_1966,
       author = {{King}, Ivan R.},
        title = "{The structure of star clusters. III. Some simple dynamical models}",
      journal = {\aj},
         year = 1966,
        month = feb,
       volume = {71},
        pages = {64},
          doi = {10.1086/109857},
       adsurl = {https://ui.adsabs.harvard.edu/abs/1966AJ.....71...64K}
}

@ARTICLE{Aarseth_2012,
  author = {{Aarseth}, S.~J.},
  title = {{Mergers and ejections of black holes in globular clusters}},
  journal = {\mnras},
  archiveprefix = {arXiv},
  eprint = {1202.4688},
  primaryclass = {astro-ph.SR},
  year = {2012},
  month = {may},
  volume = {422},
  pages = {841-848},
  doi = {10.1111/j.1365-2966.2012.20666.x},
  adsurl = {http://adsabs.harvard.edu/abs/2012MNRAS.422..841A},
}

@article{Hurley_2002,
  doi = {10.1046/j.1365-8711.2002.05038.x},
  url = {http://dx.doi.org/10.1046/j.1365-8711.2002.05038.x},
  year = {2002},
  month = {feb},
  publisher = {Oxford University Press ({OUP})},
  volume = {329},
  number = {4},
  pages = {897--928},
  author = {J. R. Hurley and C. A. Tout and O. R. Pols},
  title = {{Evolution of binary stars and the effect of tides on binary populations}},
  journal = {Monthly Notices of the Royal Astronomical Society},
}

@ARTICLE{Banerjee_2018,
   author = {{Banerjee}, S.},
    title = "{Stellar-mass black holes in young massive and open stellar clusters and their role in gravitational-wave generation III: dissecting black hole dynamics}",
  journal = {\mnras},
archivePrefix = "arXiv",
   eprint = {1805.06466},
 primaryClass = "astro-ph.HE",
     year = 2018,
    month = dec,
   volume = 481,
    pages = {5123-5145},
      doi = {10.1093/mnras/sty2608},
   adsurl = {http://adsabs.harvard.edu/abs/2018MNRAS.481.5123B}
}

@ARTICLE{Hurley_2005,
   author = {{Hurley}, J.~R. and {Pols}, O.~R. and {Aarseth}, S.~J. and {Tout}, C.~A.
	},
    title = "{A complete N-body model of the old open cluster M67}",
  journal = {\mnras},
   eprint = {astro-ph/0507239},
     year = 2005,
    month = oct,
   volume = 363,
    pages = {293-314},
      doi = {10.1111/j.1365-2966.2005.09448.x},
   adsurl = {http://adsabs.harvard.edu/abs/2005MNRAS.363..293H}
}

@BOOK{Heggie_2003,
      author = {{Heggie}, Douglas and {Hut}, Piet},
        title = "{The Gravitational Million-Body Problem: A Multidisciplinary Approach to Star Cluster Dynamics}",
         year = 2003,
       adsurl = {https://ui.adsabs.harvard.edu/abs/2003gmbp.book.....H}
}

@BOOK{Spitzer_1987,
  author = {{Spitzer}, L.},
  title = {{Dynamical evolution of globular clusters, Princeton University Press, Princeton, NJ, 191 p.}},
  booktitle = {Princeton, NJ, Princeton University Press, 1987, 191 p.},
  year = {1987},
  adsurl = {http://adsabs.harvard.edu/abs/1987degc.book.....S},
}

@ARTICLE{Kroupa_2001,
   author = {{Kroupa}, P.},
    title = "{On the variation of the initial mass function}",
  journal = {\mnras},
   eprint = {astro-ph/0009005},
     year = 2001,
    month = apr,
   volume = 322,
    pages = {231-246},
      doi = {10.1046/j.1365-8711.2001.04022.x},
   adsurl = {http://adsabs.harvard.edu/abs/2001MNRAS.322..231K}
}

@ARTICLE{PortegiesZwart_2010,
   author = {{Portegies Zwart}, S.~F. and {McMillan}, S.~L.~W. and {Gieles}, M.
	},
    title = "{Young Massive Star Clusters}",
  journal = {\araa},
archivePrefix = "arXiv",
   eprint = {1002.1961},
     year = 2010,
    month = sep,
   volume = 48,
    pages = {431-493},
      doi = {10.1146/annurev-astro-081309-130834},
   adsurl = {http://adsabs.harvard.edu/abs/2010ARA%26A..48..431P}
}

@ARTICLE{Allen_1991,
       author = {{Allen}, Christine and {Santillan}, Alfredo},
        title = "{An improved model of the galactic mass distribution for orbit computations.}",
      journal = {Rev. Mexicana Astron. Astrofis.},
         year = 1991,
        month = oct,
       volume = {22},
        pages = {255},
       adsurl = {https://ui.adsabs.harvard.edu/abs/1991RMxAA..22..255A}
}

@article{Nitadori_2012,
  doi = {10.1111/j.1365-2966.2012.21227.x},
  url = {http://dx.doi.org/10.1111/j.1365-2966.2012.21227.x},
  year = {2012},
  month = {jun},
  publisher = {Oxford University Press ({OUP})},
  volume = {424},
  number = {1},
  pages = {545--552},
  author = {Keigo Nitadori and Sverre J. Aarseth},
  title = {{Accelerating nbody6 with graphics processing units}},
  journal = {Monthly Notices of the Royal Astronomical Society},
}

@article{Mikkola_1999,
  doi = {10.1046/j.1365-8711.1999.02982.x},
  url = {http://dx.doi.org/10.1046/j.1365-8711.1999.02982.x},
  year = {1999},
  month = {dec},
  publisher = {Oxford University Press ({OUP})},
  volume = {310},
  number = {3},
  pages = {745--749},
  author = {S. Mikkola and K. Tanikawa},
  title = {{Algorithmic regularization of the few-body problem}},
  journal = {Monthly Notices of the Royal Astronomical Society},
}

@article{Mikkola_1993,
  doi = {10.1007/bf00695714},
  url = {http://dx.doi.org/10.1007/BF00695714},
  year = {1993},
  month = {nov},
  publisher = {Springer Nature},
  volume = {57},
  number = {3},
  pages = {439--459},
  author = {Seppo Mikkola and Sverre J. Aarseth},
  title = {{An implementation {ofN}-body chain regularization}},
  journal = {Celestial Mechanics {\&} Dynamical Astronomy},
}

@ARTICLE{Banerjee_2020,
       author = {{Banerjee}, S. and {Belczynski}, K. and {Fryer}, C.~L. and
         {Berczik}, P. and {Hurley}, J.~R. and {Spurzem}, R. and {Wang}, L.},
        title = "{BSE versus StarTrack: Implementations of new wind, remnant-formation, and natal-kick schemes in NBODY7 and their astrophysical consequences}",
      journal = {\aap},
         year = 2020,
        month = jul,
       volume = {639},
          eid = {A41},
        pages = {A41},
          doi = {10.1051/0004-6361/201935332},
archivePrefix = {arXiv},
       eprint = {1902.07718},
 primaryClass = {astro-ph.SR},
       adsurl = {https://ui.adsabs.harvard.edu/abs/2020A&A...639A..41B}
}

@ARTICLE{Banerjee_2020c,
    author = {{Banerjee}, Sambaran},
        title = "{Stellar-mass black holes in young massive and open stellar clusters - IV. Updated stellar-evolutionary and black hole spin models and comparisons with the LIGO-Virgo O1/O2 merger-event data}",
      journal = {\mnras},
         year = 2021,
        month = jan,
       volume = {500},
       number = {3},
        pages = {3002-3026},
          doi = {10.1093/mnras/staa2392},
archivePrefix = {arXiv},
       eprint = {2004.07382},
 primaryClass = {astro-ph.HE},
       adsurl = {https://ui.adsabs.harvard.edu/abs/2021MNRAS.500.3002B}
}

@ARTICLE{Fryer_2012,
   author = {{Fryer}, C.~L. and {Belczynski}, K. and {Wiktorowicz}, G. and 
	{Dominik}, M. and {Kalogera}, V. and {Holz}, D.~E.},
    title = "{Compact Remnant Mass Function: Dependence on the Explosion Mechanism and Metallicity}",
  journal = {\apj},
archivePrefix = "arXiv",
   eprint = {1110.1726},
 primaryClass = "astro-ph.SR",
     year = 2012,
    month = apr,
   volume = 749,
      eid = {91},
    pages = {91},
      doi = {10.1088/0004-637X/749/1/91},
   adsurl = {http://adsabs.harvard.edu/abs/2012ApJ...749...91F}
}

@article{Vink_2001,
  doi = {10.1051/0004-6361:20010127},
  url = {http://dx.doi.org/10.1051/0004-6361:20010127},
  year = {2001},
  month = {apr},
  publisher = {{EDP} Sciences},
  volume = {369},
  number = {2},
  pages = {574--588},
  author = {Jorick S. Vink and A. de Koter and H. J. G. L. M. Lamers},
  title = {{Mass-loss predictions for O and B stars as a function of metallicity}},
  journal = {Astronomy and Astrophysics},
}

@ARTICLE{Lombardi_2002,
       author = {{Lombardi}, James C., Jr. and {Warren}, Jessica S. and {Rasio}, Frederic
        A. and {Sills}, Alison and {Warren}, Aaron R.},
        title = "{Stellar Collisions and the Interior Structure of Blue Stragglers}",
      journal = {\apj},
         year = 2002,
        month = Apr,
       volume = {568},
        pages = {939-953},
          doi = {10.1086/339060},
archivePrefix = {arXiv},
       eprint = {astro-ph/0107388},
 primaryClass = {astro-ph},
       adsurl = {https://ui.adsabs.harvard.edu/\#abs/2002ApJ...568..939L}
}

@ARTICLE{Gaburov_2008,
       author = {{Gaburov}, E. and {Lombardi}, J.~C. and {Portegies Zwart}, S.},
        title = "{Mixing in massive stellar mergers}",
      journal = {\mnras},
         year = 2008,
        month = Jan,
       volume = {383},
        pages = {L5-L9},
          doi = {10.1111/j.1745-3933.2007.00399.x},
archivePrefix = {arXiv},
       eprint = {0707.3021},
 primaryClass = {astro-ph},
       adsurl = {https://ui.adsabs.harvard.edu/\#abs/2008MNRAS.383L...5G}
}

@ARTICLE{Glebbeek_2009,
       author = {{Glebbeek}, E. and {Gaburov}, E. and {de Mink}, S.~E. and {Pols}, O.~R.
        and {Portegies Zwart}, S.~F.},
        title = "{The evolution of runaway stellar collision products}",
      journal = {\aap},
         year = 2009,
        month = Apr,
       volume = {497},
        pages = {255-264},
          doi = {10.1051/0004-6361/200810425},
archivePrefix = {arXiv},
       eprint = {0902.1753},
 primaryClass = {astro-ph.SR},
       adsurl = {https://ui.adsabs.harvard.edu/\#abs/2009A&A...497..255G}
}

@ARTICLE{DeMink_2009,
   author = {{De Mink}, S.~E. and {Cantiello}, M. and {Langer}, N. and {Pols}, O.~R. and 
	{Brott}, I. and {Yoon}, S.-C.},
    title = "{Rotational mixing in massive binaries. Detached short-period systems}",
  journal = {\aap},
archivePrefix = "arXiv",
   eprint = {0902.1751},
 primaryClass = "astro-ph.SR",
     year = 2009,
    month = apr,
   volume = 497,
    pages = {243-253},
      doi = {10.1051/0004-6361/200811439},
   adsurl = {http://adsabs.harvard.edu/abs/2009A%26A...497..243D}
}

@ARTICLE{vonZeipel_1910,
       author = {{von Zeipel}, H.},
        title = "{Sur l'application des s{\'e}ries de M. Lindstedt {\`a} l'{\'e}tude du mouvement des com{\`e}tes p{\'e}riodiques}",
      journal = {Astronomische Nachrichten},
         year = 1910,
        month = mar,
       volume = {183},
       number = {22},
        pages = {345},
          doi = {10.1002/asna.19091832202},
       adsurl = {https://ui.adsabs.harvard.edu/abs/1910AN....183..345V}
}

@ARTICLE{Kozai_1962,
       author = {{Kozai}, Yoshihide},
        title = "{Secular perturbations of asteroids with high inclination and eccentricity}",
      journal = {\aj},
         year = 1962,
        month = nov,
       volume = {67},
        pages = {591-598},
          doi = {10.1086/108790},
       adsurl = {https://ui.adsabs.harvard.edu/abs/1962AJ.....67..591K}
}

@ARTICLE{Lidov_1962,
       author = {{Lidov}, M.~L.},
        title = "{The evolution of orbits of artificial satellites of planets under the action of gravitational perturbations of external bodies}",
      journal = {\planss},
         year = 1962,
        month = oct,
       volume = {9},
       number = {10},
        pages = {719-759},
          doi = {10.1016/0032-0633(62)90129-0},
       adsurl = {https://ui.adsabs.harvard.edu/abs/1962P&SS....9..719L}
}

@ARTICLE{Trani_2024,
       author = {{Trani}, Alessandro Alberto and {Leigh}, Nathan W.~C. and {Boekholt}, Tjarda C.~N. and {Portegies Zwart}, Simon},
        title = "{Isles of regularity in a sea of chaos amid the gravitational three-body problem}",
      journal = {\aap},
         year = 2024,
        month = sep,
       volume = {689},
          eid = {A24},
        pages = {A24},
          doi = {10.1051/0004-6361/202449862},
archivePrefix = {arXiv},
       eprint = {2403.03247},
 primaryClass = {astro-ph.EP},
       adsurl = {https://ui.adsabs.harvard.edu/abs/2024A&A...689A..24T}
}

@ARTICLE{Rao2026ApJ..1002..103R,
       author = {{Rao}, Khushboo K. and {Chen}, Wen Ping},
        title = "{Investigating Extended Main-sequence Turnoffs in Galactic Open Clusters}",
      journal = {\apj},
         year = 2026,
        month = may,
       volume = {1002},
       number = {1},
          eid = {103},
        pages = {103},
          doi = {10.3847/1538-4357/ae5b9e},
archivePrefix = {arXiv},
       eprint = {2604.03746},
 primaryClass = {astro-ph.GA},
       adsurl = {https://ui.adsabs.harvard.edu/abs/2026ApJ..1002..103R}
}

@ARTICLE{Li2012ApJ...761L..22L,
       author = {{Li}, Zhongmu and {Mao}, Caiyan and {Chen}, Li and {Zhang}, Qian},
        title = "{Combined Effects of Binaries and Stellar Rotation on the Color-Magnitude Diagrams of Intermediate-age Star Clusters}",
      journal = {\apjl},
         year = 2012,
        month = dec,
       volume = {761},
       number = {2},
          eid = {L22},
        pages = {L22},
          doi = {10.1088/2041-8205/761/2/L22},
archivePrefix = {arXiv},
       eprint = {1302.0099},
 primaryClass = {astro-ph.SR},
       adsurl = {https://ui.adsabs.harvard.edu/abs/2012ApJ...761L..22L}
}

@ARTICLE{Correnti2021MNRAS.504..155C,
       author = {{Correnti}, Matteo and {Goudfrooij}, Paul and {Bellini}, Andrea and {Girardi}, Leo},
        title = "{The wide upper main sequence and main-sequence turnoff of the {\ensuremath{\sim}} 800 Myr old star cluster NGC 1831}",
      journal = {\mnras},
         year = 2021,
        month = jun,
       volume = {504},
       number = {1},
        pages = {155-165},
          doi = {10.1093/mnras/stab233},
archivePrefix = {arXiv},
       eprint = {2101.10751},
 primaryClass = {astro-ph.GA},
       adsurl = {https://ui.adsabs.harvard.edu/abs/2021MNRAS.504..155C}
}

@ARTICLE{Bastian2017MNRAS.465.4795B,
       author = {{Bastian}, N. and {Cabrera-Ziri}, I. and {Niederhofer}, F. and {de Mink}, S.~E. and {Georgy}, C. and {Baade}, D. and {Correnti}, M. and {Usher}, C. and {Romaniello}, M.},
        title = "{A high fraction of Be stars in young massive clusters: evidence for a large population of near-critically rotating stars}",
      journal = {\mnras},
         year = 2017,
        month = mar,
       volume = {465},
       number = {4},
        pages = {4795-4799},
          doi = {10.1093/mnras/stw3042},
archivePrefix = {arXiv},
       eprint = {1611.06705},
 primaryClass = {astro-ph.GA},
       adsurl = {https://ui.adsabs.harvard.edu/abs/2017MNRAS.465.4795B}
}

@ARTICLE{Leanza2025A&A...698A..27L,
       author = {{Leanza}, S. and {Dalessandro}, E. and {Cadelano}, M. and {Fanelli}, C. and {Ettorre}, G. and {Kamann}, S. and {Bastian}, N. and {Martocchia}, S. and {Salaris}, M. and {Lardo}, C. and {Mucciarelli}, A. and {Saracino}, S.},
        title = "{Stellar rotation in the intermediate-age massive cluster NGC 1783: Clues about the nature of UV-dim stars}",
      journal = {\aap},
         year = 2025,
        month = jun,
       volume = {698},
          eid = {A27},
        pages = {A27},
          doi = {10.1051/0004-6361/202553956},
archivePrefix = {arXiv},
       eprint = {2504.08362},
 primaryClass = {astro-ph.GA},
       adsurl = {https://ui.adsabs.harvard.edu/abs/2025A&A...698A..27L}
}

@ARTICLE{Rao2025arXiv251205458R,
       author = {{Rao}, Khushboo K and {Chen}, Wen-Ping},
        title = "{Investigating Unusually Extended Main Sequence Turnoff of the Galactic Open Cluster NGC 3532}",
      journal = {arXiv e-prints},
         year = 2025,
        month = dec,
          eid = {arXiv:2512.05458},
        pages = {arXiv:2512.05458},
          doi = {10.48550/arXiv.2512.05458},
archivePrefix = {arXiv},
       eprint = {2512.05458},
 primaryClass = {astro-ph.GA},
       adsurl = {https://ui.adsabs.harvard.edu/abs/2025arXiv251205458R}
}

@ARTICLE{Nguyen2022A&A...665A.126N,
       author = {{Nguyen}, C.~T. and {Costa}, G. and {Girardi}, L. and {Volpato}, G. and {Bressan}, A. and {Chen}, Y. and {Marigo}, P. and {Fu}, X. and {Goudfrooij}, P.},
        title = "{PARSEC V2.0: Stellar tracks and isochrones of low- and intermediate-mass stars with rotation}",
      journal = {\aap},
         year = 2022,
        month = sep,
       volume = {665},
          eid = {A126},
        pages = {A126},
          doi = {10.1051/0004-6361/202244166},
archivePrefix = {arXiv},
       eprint = {2207.08642},
 primaryClass = {astro-ph.SR},
       adsurl = {https://ui.adsabs.harvard.edu/abs/2022A&A...665A.126N}
}

@ARTICLE{Clem2011AJ....141..115C,
       author = {{Clem}, James L. and {Landolt}, Arlo U. and {Hoard}, D.~W. and {Wachter}, Stefanie},
        title = "{Deep, Wide-field CCD Photometry for the Open Cluster NGC 3532}",
      journal = {\aj},
         year = 2011,
        month = apr,
       volume = {141},
       number = {4},
          eid = {115},
        pages = {115},
          doi = {10.1088/0004-6256/141/4/115},
archivePrefix = {arXiv},
       eprint = {1101.3268},
 primaryClass = {astro-ph.SR},
       adsurl = {https://ui.adsabs.harvard.edu/abs/2011AJ....141..115C}
}

@ARTICLE{Cordoni2023A&A...672A..29C,
       author = {{Cordoni}, Giacomo and {Milone}, Antonino P. and {Marino}, Anna F. and {Vesperini}, Enrico and {Dondoglio}, Emanuele and {Legnardi}, Maria Vittoria and {Mohandasan}, Anjana and {Carlos}, Marilia and {Lagioia}, Edoardo P. and {Jang}, Sohee and {Ziliotto}, Tuila},
        title = "{Photometric binaries, mass functions, and structural parameters of 78 Galactic open clusters}",
      journal = {\aap},
         year = 2023,
        month = apr,
       volume = {672},
          eid = {A29},
        pages = {A29},
          doi = {10.1051/0004-6361/202245457},
archivePrefix = {arXiv},
       eprint = {2302.03685},
 primaryClass = {astro-ph.SR},
       adsurl = {https://ui.adsabs.harvard.edu/abs/2023A&A...672A..29C}
}

@ARTICLE{Ginsburg2019AJ....157...98G,
       author = {{Ginsburg}, Adam and {Sip{\H{o}}cz}, Brigitta M. and {Brasseur}, C.~E. and {Cowperthwaite}, Philip S. and {Craig}, Matthew W. and {Deil}, Christoph and {Guillochon}, James and {Guzman}, Giannina and {Liedtke}, Simon and {Lian Lim}, Pey and {Lockhart}, Kelly E. and {Mommert}, Michael and {Morris}, Brett M. and {Norman}, Henrik and {Parikh}, Madhura and {Persson}, Magnus V. and {Robitaille}, Thomas P. and {Segovia}, Juan-Carlos and {Singer}, Leo P. and {Tollerud}, Erik J. and {de Val-Borro}, Miguel and {Valtchanov}, Ivan and {Woillez}, Julien and {Astroquery Collaboration} and {a subset of astropy Collaboration}},
        title = "{astroquery: An Astronomical Web-querying Package in Python}",
      journal = {\aj},
         year = 2019,
        month = mar,
       volume = {157},
       number = {3},
          eid = {98},
        pages = {98},
          doi = {10.3847/1538-3881/aafc33},
archivePrefix = {arXiv},
       eprint = {1901.04520},
 primaryClass = {astro-ph.IM},
       adsurl = {https://ui.adsabs.harvard.edu/abs/2019AJ....157...98G}
}

@Article{Hunter:2007,
  Author    = {Hunter, J. D.},
  Title     = {Matplotlib: A 2D graphics environment},
  Journal   = {Computing in Science \& Engineering},
  Volume    = {9},
  Number    = {3},
  Pages     = {90--95},
  publisher = {IEEE COMPUTER SOC},
  doi       = {10.1109/MCSE.2007.55},
  year      = 2007
}

@ARTICLE{2020Natur.585..357Harris,
       author = {{Harris}, Charles R. and {Millman}, K. Jarrod and {van der Walt}, St{\'e}fan J. and {Gommers}, Ralf and {Virtanen}, Pauli and {Cournapeau}, David and {Wieser}, Eric and {Taylor}, Julian and {Berg}, Sebastian and {Smith}, Nathaniel J. and {Kern}, Robert and {Picus}, Matti and {Hoyer}, Stephan and {van Kerkwijk}, Marten H. and {Brett}, Matthew and {Haldane}, Allan and {del R{\'\i}o}, Jaime Fern{\'a}ndez and {Wiebe}, Mark and {Peterson}, Pearu and {G{\'e}rard-Marchant}, Pierre and {Sheppard}, Kevin and {Reddy}, Tyler and {Weckesser}, Warren and {Abbasi}, Hameer and {Gohlke}, Christoph and {Oliphant}, Travis E.},
        title = "{Array programming with NumPy}",
      journal = {\nat},
         year = 2020,
        month = sep,
       volume = {585},
       number = {7825},
        pages = {357-362},
          doi = {10.1038/s41586-020-2649-2},
archivePrefix = {arXiv},
       eprint = {2006.10256},
 primaryClass = {cs.MS},
       adsurl = {https://ui.adsabs.harvard.edu/abs/2020Natur.585..357H}
}

@PROCEEDINGS{Boffin2015ASSL..413.....B,
        title = "{Ecology of Blue Straggler Stars}",
    booktitle = {Astrophysics and Space Science Library},
         year = 2015,
       editor = {{Boffin}, Henri M.~J. and {Carraro}, Giovanni and {Beccari}, Giacomo},
       series = {Astrophysics and Space Science Library},
       volume = {413},
        month = jan,
          doi = {10.1007/978-3-662-44434-4},
archivePrefix = {arXiv},
       eprint = {1406.3909},
 primaryClass = {astro-ph.SR},
       adsurl = {https://ui.adsabs.harvard.edu/abs/2015ASSL..413.....B}
}

@ARTICLE{Ferraro2023NatCo..14.2584F,
       author = {{Ferraro}, Francesco R. and {Mucciarelli}, Alessio and {Lanzoni}, Barbara and {Pallanca}, Cristina and {Cadelano}, Mario and {Billi}, Alex and {Sills}, Alison and {Vesperini}, Enrico and {Dalessandro}, Emanuele and {Beccari}, Giacomo and {Monaco}, Lorenzo and {Mateo}, Mario},
        title = "{Fast rotating blue stragglers prefer loose clusters}",
      journal = {Nature Communications},
         year = 2023,
        month = may,
       volume = {14},
          eid = {2584},
        pages = {2584},
          doi = {10.1038/s41467-023-38153-w},
archivePrefix = {arXiv},
       eprint = {2305.08478},
 primaryClass = {astro-ph.SR},
       adsurl = {https://ui.adsabs.harvard.edu/abs/2023NatCo..14.2584F}
}

@ARTICLE{Rao2023MNRAS.526.1057R,
       author = {{Rao}, Khushboo K. and {Vaidya}, Kaushar and {Agarwal}, Manan and {Balan}, Shanmugha and {Bhattacharya}, Souradeep},
        title = "{Determination of dynamical ages of open clusters through the A$^{+}$ parameter - II}",
      journal = {\mnras},
         year = 2023,
        month = nov,
       volume = {526},
       number = {1},
        pages = {1057-1074},
          doi = {10.1093/mnras/stad2755},
archivePrefix = {arXiv},
       eprint = {2309.02746},
 primaryClass = {astro-ph.GA},
       adsurl = {https://ui.adsabs.harvard.edu/abs/2023MNRAS.526.1057R}
}

@ARTICLE{Sandage1953AJ.....58...61S,
       author = {{Sandage}, A.~R.},
        title = "{The color-magnitude diagram for the globular cluster M 3.}",
      journal = {\aj},
         year = 1953,
        month = jan,
       volume = {58},
        pages = {61-75},
          doi = {10.1086/106822},
       adsurl = {https://ui.adsabs.harvard.edu/abs/1953AJ.....58...61S}
}

@ARTICLE{Naoz2016ARA&A..54..441N,
       author = {{Naoz}, Smadar},
        title = "{The Eccentric Kozai-Lidov Effect and Its Applications}",
      journal = {\araa},
         year = 2016,
        month = sep,
       volume = {54},
        pages = {441-489},
          doi = {10.1146/annurev-astro-081915-023315},
archivePrefix = {arXiv},
       eprint = {1601.07175},
 primaryClass = {astro-ph.EP},
       adsurl = {https://ui.adsabs.harvard.edu/abs/2016ARA&A..54..441N}
}

@ARTICLE{Henneco2024A&A...682A.169H,
       author = {{Henneco}, J. and {Schneider}, F.~R.~N. and {Laplace}, E.},
        title = "{Contact tracing of binary stars: Pathways to stellar mergers}",
      journal = {\aap},
         year = 2024,
        month = feb,
       volume = {682},
          eid = {A169},
        pages = {A169},
          doi = {10.1051/0004-6361/202347893},
archivePrefix = {arXiv},
       eprint = {2311.12124},
 primaryClass = {astro-ph.SR},
       adsurl = {https://ui.adsabs.harvard.edu/abs/2024A&A...682A.169H}
}

@ARTICLE{Bastian2025A&A...700A.241B,
       author = {{Bastian}, N. and {Kamann}, S. and {Niederhofer}, F. and {Saracino}, S.},
        title = "{Testing the role of merging binaries in the formation of the split main sequence in young clusters}",
      journal = {\aap},
         year = 2025,
        month = aug,
       volume = {700},
          eid = {A241},
        pages = {A241},
          doi = {10.1051/0004-6361/202555369},
archivePrefix = {arXiv},
       eprint = {2509.07708},
 primaryClass = {astro-ph.SR},
       adsurl = {https://ui.adsabs.harvard.edu/abs/2025A&A...700A.241B}
}
%%%%%%%%%%%%%%%%%%%%%%%%%%%%%%%%%%%%%%%%%%%%%%%%%%%%%%%%%%%%%%%
\begin{appendix}

\section{Determination of structural parameters }\label{sec:structural_params}

The Gaia EDR3 data are 90\% complete down to a magnitude of $G=19.5$. To address the issue of incompleteness, we select cluster members only down to $G=18.5$ for estimating the structural parameters. We begin by estimating the number density of cluster members in various annular regions of equal radius. Using the LIMPY Python package, we then apply emcee, an affine-invariant Markov Chain Monte Carlo (MCMC) sampler, to estimate the parameters and uncertainties of the King model \citep{King_1966}. We initially vary the number of bins from 50 to 100 and fit the single-mass isotropic King model. The best-fitting number density profile is visually determined using the minimum reduced $\chi^2$ method. To obtain the initial set of parameters, we employ the differential evolution global optimization method. These initial parameters, along with their ranges, are then provided to emcee for the estimation of final parameters and their errors. We then run emcee for 100 walkers, 100 burn-in steps, and 4000 iterations. The final structural parameters and mass of the cluster are: $r_{\rm c} = 15.62 \pm 0.66$~arcmin, $r_{\rm h} = 42.33 \pm 4.38$~arcmin, $r_{\rm t} = 292.94 \pm 53.74$~arcmin, $M_{\rm cl} = 2545 \pm 218$~M$_\odot$. The King model-fitted number density profile of the cluster is shown in Fig.\ref{fig:king_profile}.

%fig app 01
\begin{figure}
\begin{center}
    \includegraphics[width=1.0\linewidth]{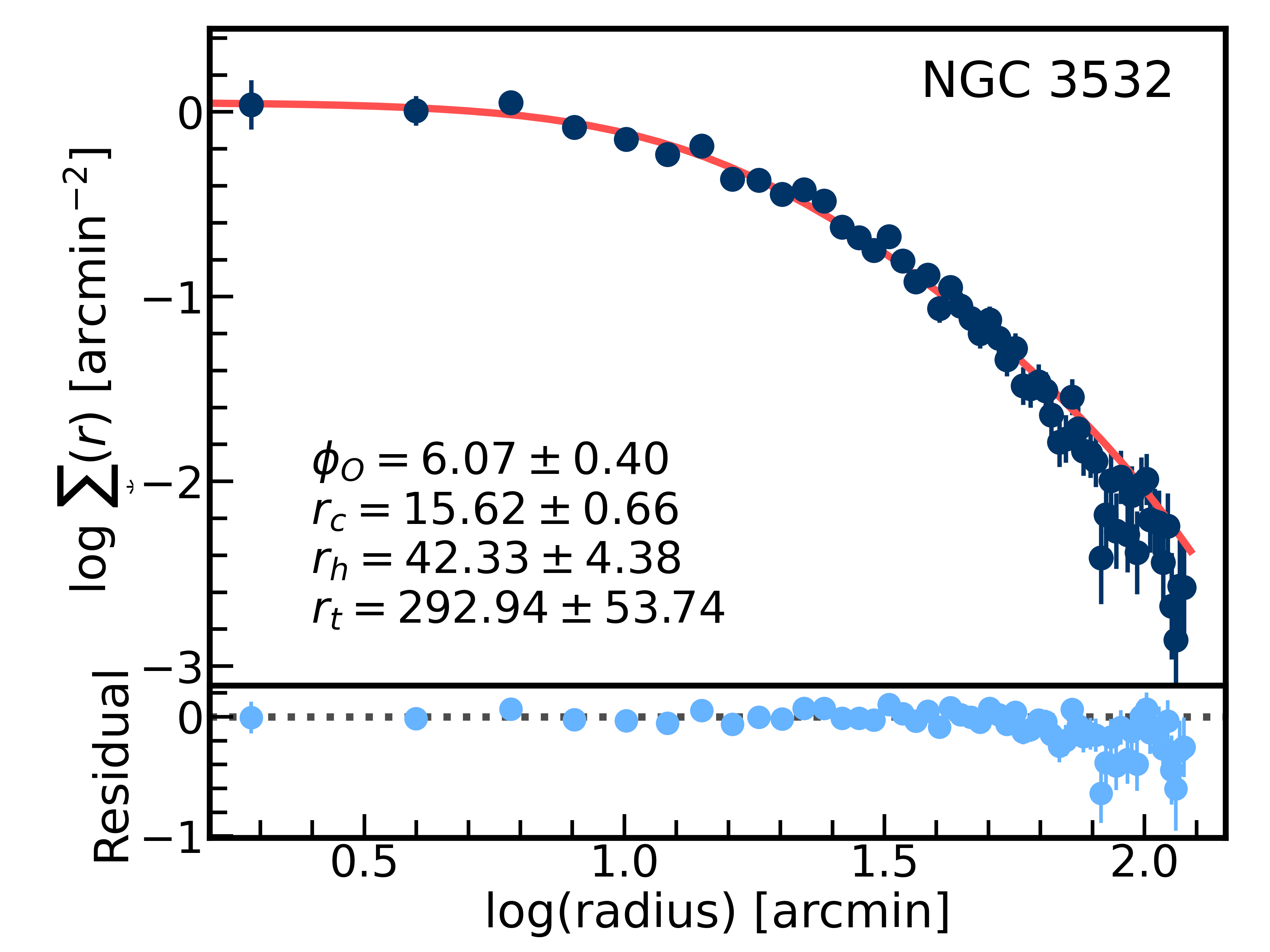}
\caption{Observed radial number-density profile of NGC\,3532 fitted with a single-mass, isotropic King model \citep{King_1966} (upper panel). The lower panel shows the residuals of the fit.}
\label{fig:king_profile}
\end{center}
\end{figure}

%fig app 02
\begin{figure}
\begin{center}
\includegraphics[width=0.95\linewidth]{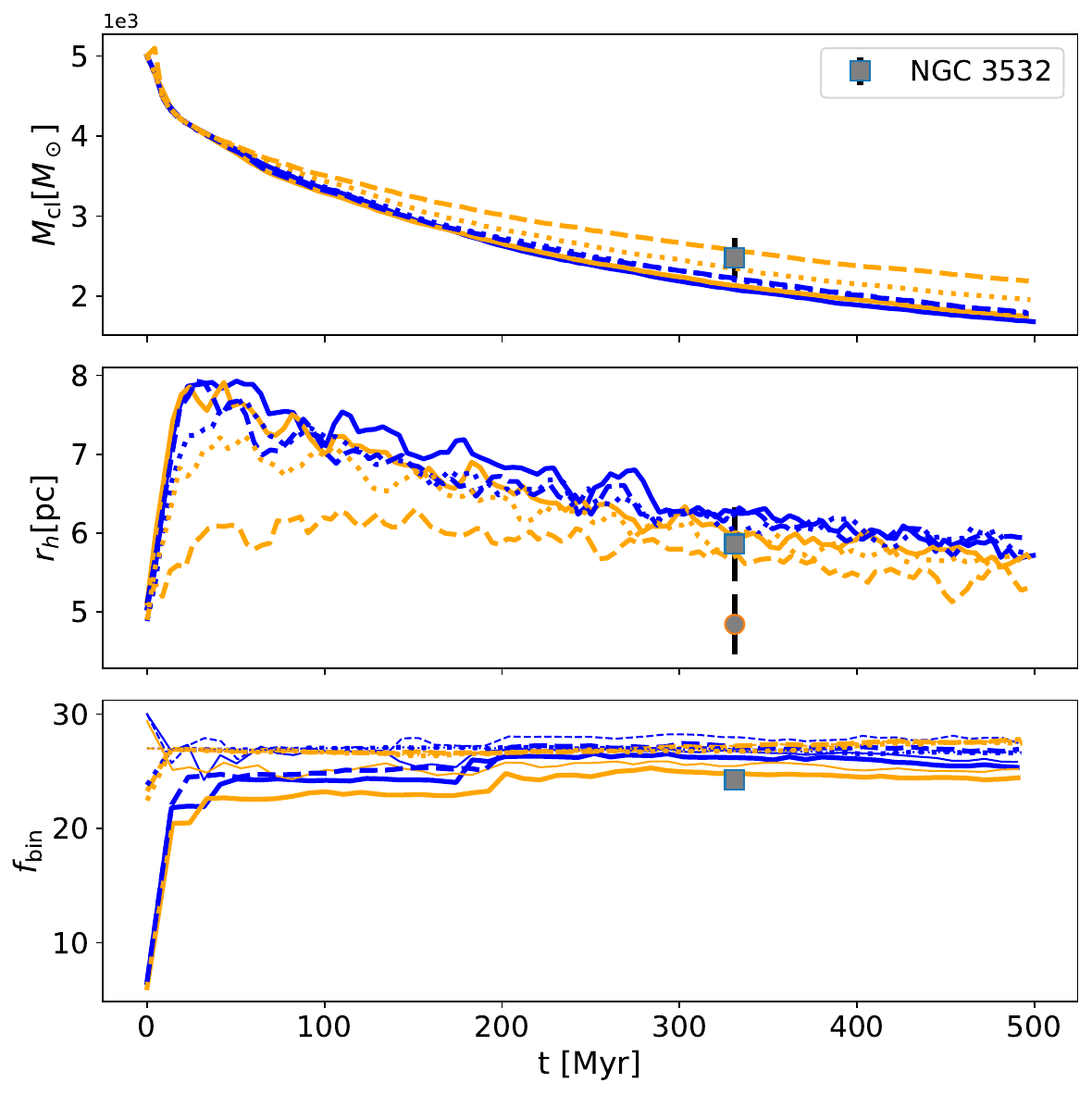}
	\caption{Same as Fig.~\ref{fig:Nbevol1}, but for the model star clusters with
	primordial binaries having a flat initial semi-major-axis distribution (see Table~\ref{tab:NGC3532_models}).}
\label{fig:Nbevol2}
\end{center}
\end{figure}

\end{appendix}
\end{document}